\documentclass{jkas}

\def\year{2018} 
\def\month{April} 
\def\volume{00} 
\def\issue{0} 
\def\beginpage{1} 
\def\endpage{99} 
\def\accepted{---} 
\date{received; September 2020 ; accepted \accepted}

\newcommand{\dego}{$^{\circ}$}

\usepackage{graphicx,times}          
\usepackage{natbib}
\usepackage{amssymb,amsmath}
\usepackage[a4paper=true,dvipdfm=true,pagebackref=true]{hyperref}

\title{
Quadruply-imaged quasars: some general features
}

\author[1]{P. Tuan-Anh}
\author[1]{T. T. Thai}
\author[2]{N. A. Tuan}
\author[1]{P. Darriulat}
\author[1]{P. N. Diep}
\author[1]{D. T. Hoai}
\author[1]{N. B. Ngoc}
\author[1]{P. T. Nhung}
\author[1]{N. T. Phuong}

\affil[1]{Department of Astrophysics, Vietnam National Space Center, VAST, 18 Hoang Quoc Viet, Hanoi, Vietnam; \email{ptanh@vnsc.org.vn}}
\affil[2]{University of Science and Technology of Hanoi (USTH), VAST, 18 Hoang Quoc Viet, Hanoi, Vietnam; }

\def\corrauthor{
P. Tuan-Anh
}

\def\runningauthor{
P. Tuan-Anh et al.
}

\def\runningtitle{
Quadruply-imaged quasars
}

\def\keywords{
 Gravitational lensing, quasars
}

\def\abstracttext{
Gravitational  lensing  of  point  sources  located  inside  the  lens  caustic  is  known to  produce  four  images  in  a configuration closely related to the source position. We study this relation in the particular case of a sample of quadruply-imaged quasars observed by the Hubble Space Telescope (HST). Strong correlations between the  parameters defining  the  image  configuration  are  revealed. The relation between the image configuration and the source position is studied. Some  simple  features  of  the selected  data  sample  are  exposed  and  commented upon. In particular, evidence is found for the selected sample to be biased in favour of large magnification systems. While having no direct impact on practical analyses of specific systems, the results have pedagogical value and deepen our understanding of the mechanism of gravitational lensing.
}

\begin{document}
\jkashead 

\section{INTRODUCTION}\label{sec1}
\begin{figure*}
  \centering
    \includegraphics[width=1.\textwidth,trim=0.cm 1.5cm 0.cm 2.2cm,clip]{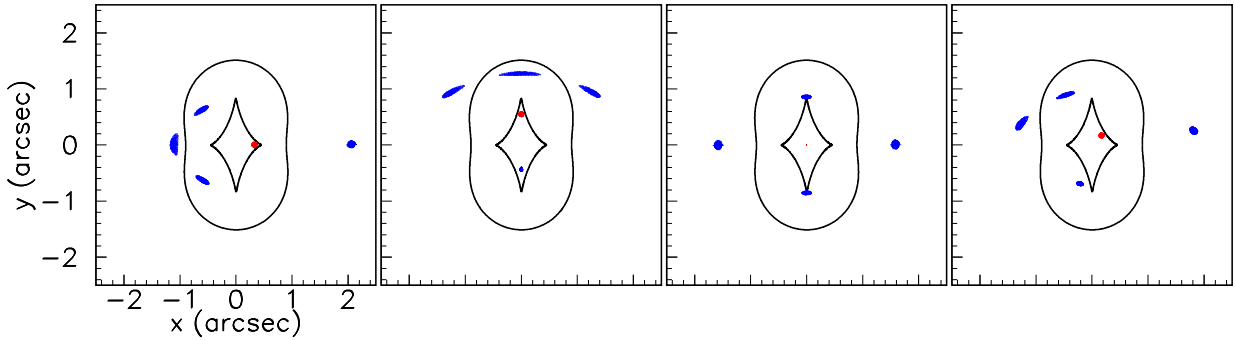}
\caption{Typical quadruply-imaged  quasars  according  to \citet*{Saha2003}  classification. Both images and sources are shown on the plane of the sky ($x,y$).  The images (blue areas) are of a disc source (red disc) having a radius of 0.05 arcsec; the caustic and critical curves are shown as black lines (from \citealt{Hoai2013}). From  left  to right: short axis quad, long axis quad, core quad and inclined quad.}
\label{fig1}
\end{figure*}

 Strong  gravitational  lensing  is  an  important  tool  for  astrophysics  observations,  in  particular  by amplifying  the  observed  brightness  of  sources  of  the  early  Universe  and  thereby  allowing  for  reaching farther  out  than  would  otherwise  be  possible.  Numerous  textbooks  and  journal  articles  are  dedicated  to  its study. The present article addresses the case of a specific sample of gravitationally lensed quasars, those for which  four  separated  images  have  been  observed. Systematic  studies  of  such  systems,  called  quads  in  the gravitational  lensing  jargon,  are  available  in  the  literature. Many of these study the case of a nearly isotropic lensing potential, the anisotropy being described to first order only. Indeed, in most practical real cases, such a treatment is known to provide a good description of the image configuration. In many cases, the anisotropy is due to a slightly aspherical mass distribution of the lensing galaxy, in which case it is natural to describe it with an ellipticity term; in many other cases it is due to the tidal effect of other galaxies in the neighbourhood of the lens, in which case it is best described by a quadrupole tidal shear term. Both descriptions give similar results but \citet{Kovner1987} was first to underline the outstanding simplicity of the latter description. He remarked that independently from the orientation and scale of the images, their configuration depends on a single parameter, the tidal shear, and that the lens equation reduces to a single equation having this parameter as unknown. He gave recipes to solve it graphically and stated some general rules relating the image multiplicity to the position of the source in the lens caustic. A few years later, together with Kassiola \citep{Kassiola1995} he used the same simple lensing potential to construct two parameters allowing for an educated guess to be made of whether the lens has a simple mass distribution. Meanwhile, together with Blandford and Kochanek \citep{Blandford1989} he had produced a first comprehensive review of gravitational lensing, a field of astrophysics that was then developing rapidly. In 2003, in the same spirit of simplicity, \mbox{\citet{Saha2003}} remarked that some characteristics of multiply imaged QSO systems are very model independent and can be deduced accurately by simply scrutinizing the image configuration; depending on the position of the source relative to the caustic they identified four kinds of quadruple systems which they named core quads, inclined quads, long-axis quads and short-axis quads. An illustration is given in Figure \ref{fig1}. Other names have been later used by other authors, such as crosses, folds and cusps \citep{Keeton2003, Keeton2005}, the latter class including both the long-axis and short-axis quads of the Saha \& Williams classification; indeed, the distinction between the two is only possible when the lensing potential is significantly aspherical. A decade later, again in the same spirit of obtaining simply useful information about the lensing mechanism without any recourse to mass modelling, \mbox{\citet{Woldesenbet2012}} have shown that the three opening angles defining the image configuration are strongly correlated, a result that remains valid when the tidal shear or ellipticity term takes relatively large values. They investigate how well this correlation is obeyed by observed quads and discuss cases where deviations are significant, such as for RX J0911.  Another form of the correlation obeyed by quads was found by \citet{Witt1996}: the four images, the quasar and the lens are located on a same rectangular hyperbola whose asymptotes align with the potential major and minor axes; last year, \citet{Schechter2019} examined how well this correlation is obeyed by observed quads. Our work is in the legacy of these authors, particularly of that of \mbox{\citet{Woldesenbet2012}}, extending their analysis to radial distances as well as opening angles and basing it on the quasar images exclusively without requiring the lens to be observed.

Some  two  decades  ago, a  collaboration  between  scientists  at  the  Harvard-Smithsonian  Center  for Astrophysics  and  the  University  of  Arizona  \citep{Falco1999}  took  the  initiative  to  carry  out  a  complete survey  of  all  known  galaxy-mass  gravitational  lens  systems  with  image  separations  of  less  than  10 arcseconds. The survey, named CASTLeS for CfA-Arizona Space Telescope Lens Survey, used the Hubble Space Telescope (HST) to obtain deep, high-resolution images in the optical and near infrared. Since then, the collaboration continues to maintain a web site (\cite{castles}) updated with the latest HST observations of lens systems. It constitutes a very precious data base that is used by many authors. The present article studies a  subset  of  the  \mbox{\cite{castles}}  data  base  made  of  quadruply-imaged  quasars  with  the  aim  of offering a new illustration of the remarkable properties of gravitational lensing optics.
\section{THE DATA SAMPLE}  \label{sec2}
\subsection{Normalised coordinates} \label{sec2.1}
We select from the list of quadruply-imaged systems (quads) compiled by \citet{castles} those for which the  identity  of the  four  images  and  of  the  lens  seem  to  be  reliably  assessed.  This  leaves  us  with  23  cases listed  in  Table  \ref{tab1}.  In  a  first  step  we  produce,  for  each  quad,  parameters  that  define  its  configuration independently  from  its  location,  orientation  and  size. As the lens is often more difficult to identify than the images, we ignore it in our definition of normalised coordinates, which does not require the lens to be detected.  Precisely,  we  choose  coordinates  in  the  sky  plane, $x$=$r\cos\theta$ and $y$=$r\sin\theta$, such that for each quad $\sum_{1,4} x_i$=$\sum_{1,4}y_i$=0. The resulting values of $r_i$=$(x_i^2+y_i^2)^{1/2}$ provide a measure of the extension of the quad, $R_0$=$0.5\{\sum_{1,4} r_i^2\}^{1/2}$, listed in Table \ref{tab1}. Their list shows an outstanding case, that of \mbox{SDSS J1004+4112}, with $R_0$=7.4 arcsec compared with mean$\pm$rms values of 0.9$\pm$0.4 arcsec for the rest of the sample. It is a quasar at redshift $z_S\sim$1.7, lensed by a cluster of galaxies at $z_L\sim$0.7, that has been described by \mbox{\citet{Oguri2004}}. Taking $R_0$ as unit of angular separation and choosing the orientation of the axes such  that  the  image  having  the  largest  value  of $r$ is  on  the  positive  part  of  the $x$ axis,  leaves  us  with eight image coordinates, $(x_i, y_i)_{i=1,4}$, obeying four relations:
\begin{equation*}
\sum_{1,4}x_i=0; \hspace{0.2cm} \sum_{1,4}y_i=0; \hspace{0.2cm} \sum_{1,4}r_i^2=4;\hspace{0.2cm} y_1=0 \hspace{0.2 cm} (x_1 >0)  \label{1}
\end{equation*}
\begin{table*}
  \centering
  \caption{The \citet{castles} quad sample. Here $r_{ij}$ and $\theta_{ij}$ are defined in Section \ref{sec2.2} together with $r^*_{ij}$ and $\theta^*_{ij}$; $x_L$ and $y_L$ are normalised lens coordinates that are studied in Section \ref{sec3.4}.}
  \label{tab1}
  \begin{tabular}{c c c c c c c c c c}
    \hline\hline
    \textit{Nr} &
    \textit{Name}&
    $R_0 (\textit{mas})$&
    $r_{13}/r_{13}^*(\%)$&
    $\theta_{24}$ / $\theta^*_{24}(\text{\dego})$&
    $r_{24}/r_{24}^*(\%)$&
    $\theta_{13}/\theta^*_{13} (\text{\dego})$&
    $x_L (\%)$&$y_L(\%)$&
    $\delta_{13}/\delta_{24}$\\
    \hline
    1
    &{PMNJ0134-0931}
    &332
    &85/78
    &241/248
    &12/13
    &195/194
    &$-$49
    &$-$1
    &1.2/0.2\\
    
    2
    &{HE0230-2130}
    &988
    &18/23
    &200/196
    &23/20
    &197/201
    &$-$11
    &37
    &0.7/0.7\\
  
    3
    &{MG0414+0534}
    &1116
    &54/53
    &223/224
    &$-$35/$-$37
    &145/146
    &40
    &$-$12
    &0.3/0.3\\
    
    4
    &{HE0435-1223}
    &1201
    &2/5
    &182/179
    &$-$19/$-$16
    &170/167
    &0
    &$-$14
    &0.4/0.7\\
  
    5
    &{B0712+472}
    &642
    &63/63
    &233/233
    &$-$19/$-$27
    &148/156
    &58
    &1
    &0.0/1.6\\
    6
    &{HS0810+2554}
    &463
    &53/52
    &222/223
    &$-$29/$-$31
    &150/152
    &28
    &$-$7
    &0.2/0.4\\
 
    7
    &{RXJ0911+0551}
    &1377
    &98/110
    &289/278
    &6/1
    &178/183
    &1
    &2
    &2.0/1.0\\
   
    8
    &{SDSS0924-0219}
    &863
    &41/39
    &209/211
    &$-$21/$-$19
    &165/163
    &14
    &$-$12
    &0.4/0.3\\
 
    9
    &{SDSS1004+4112}
    &7412
    &45/45
    &217/217
    &$-$46/$-$41
    &147/142
    &3
    &$-$45
    &0.0/1.1\\
  
    10
    &{SDSS1011+0143}
    &1845
    &7/9
    &184/183
    &15/11
    &188/192
    &$-$1
    &2
    &0.3/0.9\\
   
    11
    &{PG1115+080}
    &1146
    &49/50
    &223/222
    &20/22
    &204/203
    &12
    &17
    &0.2/0.3\\
   
    12
    &{RXJ1131-1231}
    &1565
    &86/78
    &240/247
    &2/0
    &180/182
    &66
    &$-$1
    &1.3/0.4\\
   
    13
    &{SDSS1138+0314}
    &668
    &19/19
    &193/193
    &$-$21/$-$18
    &168/164
    &4
    &$-$15
    &0.1/0.7\\
    
    14
    &{HST12531-2914}
    &613
    &7/8
    &184/183
    &$-$26/$-$21
    &167/162
    &$-$2
    &$-$17
    &0.2/1.1\\
  
    15
    &{HST14113+5211}
    &925
    &8/11
    &187/185
    &8/5
    &184/187
    &2
    &14
    &0.5/0.5\\
   
    16
    &{H1413+117}
    &617
    &2/5
    &182/180
    &$-$15/$-$12
    &172/170
    &6
    &$-$12
    &0.5/0.5\\
   
    17
    &{HST14176+5226}
    &1430
    &8/10
    &186/184
    &$-$12/$-$10
    &173/172
    &$-$4
    &$-$10
    &0.4/0.4\\
    
    18
    &{B1422+231}
    &677
    &68/58
    &221/229
    &$-$14/$-$16
    &164/166
    &89
    &0
   & 1.5/0.5\\
   
    19
    &{B1555+375}
    &226
    &53/51
    &222/223
    &$-$32/$-$34
    &147/149
    &26
    &$-$5
    &0.2/0.4\\
   
    20
    &{WF12026-4536}
    &658
    &59/58
    &227/229
    &12/14
    &198/195
    &19
    &11
    &0.2/0.5\\
    
    21
    &{WF12033-4723}
    &1148
    &49/49
    &220/220
    &$-$27/$-$25
    &160/158
    &10
    &$-$22
    &0.0/0.5\\
    
    22
    &{B2045+265}
    &872
    &104/106
    &275/274
    &$-$7/$-$10
    &169/172
    &98
    &0
    &0.3/0.6\\
   
    23
    &{Q2237+030}
    &877
    &7/9
    &185/183
    &$-$24/$-$20
    &167/163
    &0
    &$-$20
    &0.3/0.8\\
    \hline\hline
  \end{tabular}
\end{table*}
\subsection{Correlations}\label{sec2.2}
The  definition  of  normalised  image  coordinates  described  in  the  preceding  paragraph  leaves  four parameters  defining  the  quad  configuration  independently  from  location,  orientation  and  size. We find it convenient to express these in terms of the image polar coordinates, $r_i$ and $\theta_i$=$\tan^{-1}(y_i/x_i)$. We rank the images in order of increasing values of $\theta_i$, which we choose to define between 0 and 360\dego($\theta_1$=0).
\begin{figure*}
    \includegraphics[width=0.3\textwidth,trim=.5cm 1.5cm 1.2cm 2.cm,clip]{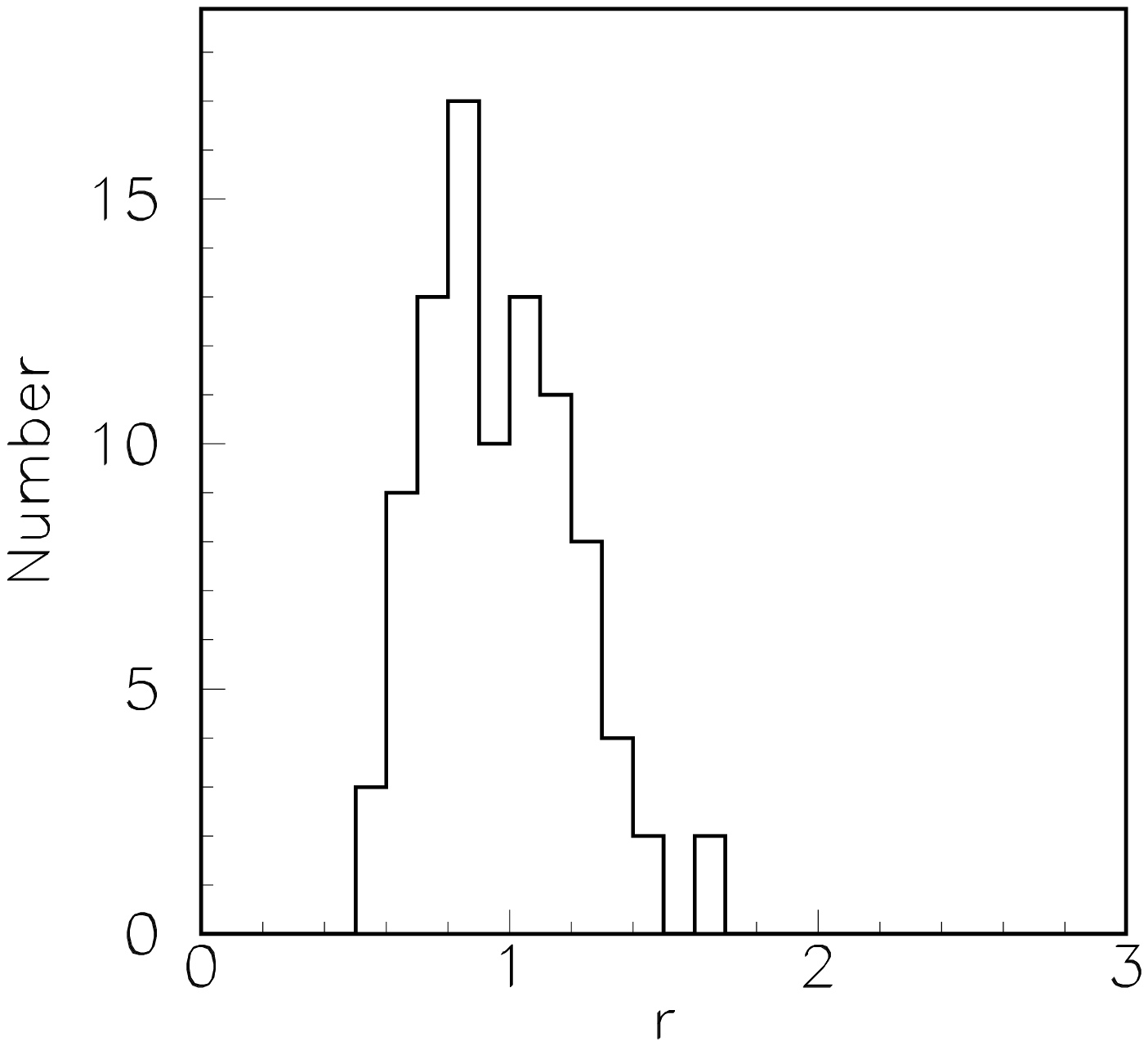}
    \includegraphics[width=0.3\textwidth,trim=.5cm 1.5cm 1.2cm 2.cm,clip]{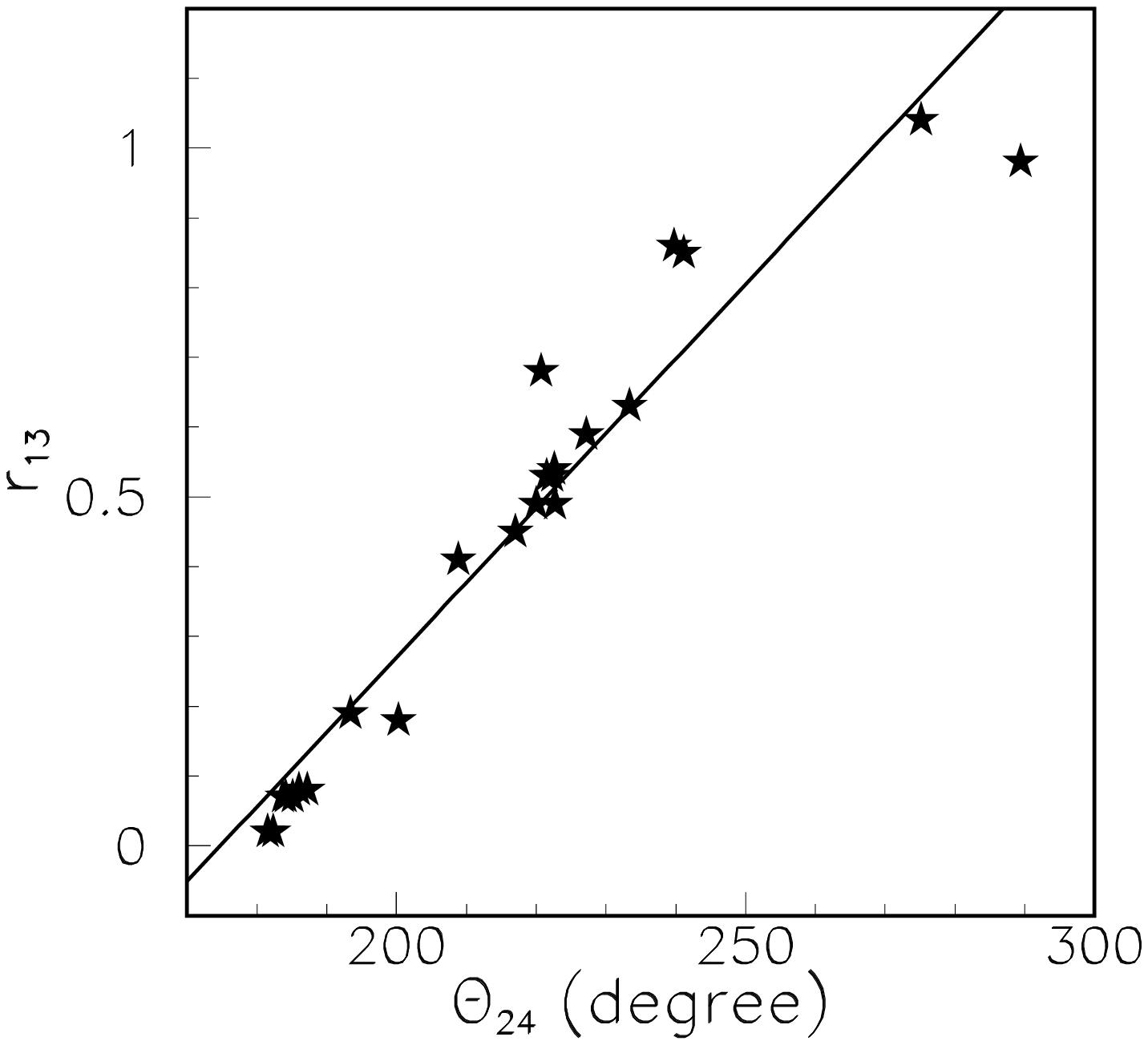}
    \includegraphics[width=0.3\textwidth,trim=.5cm 1.5cm 1.2cm 2.cm,clip]{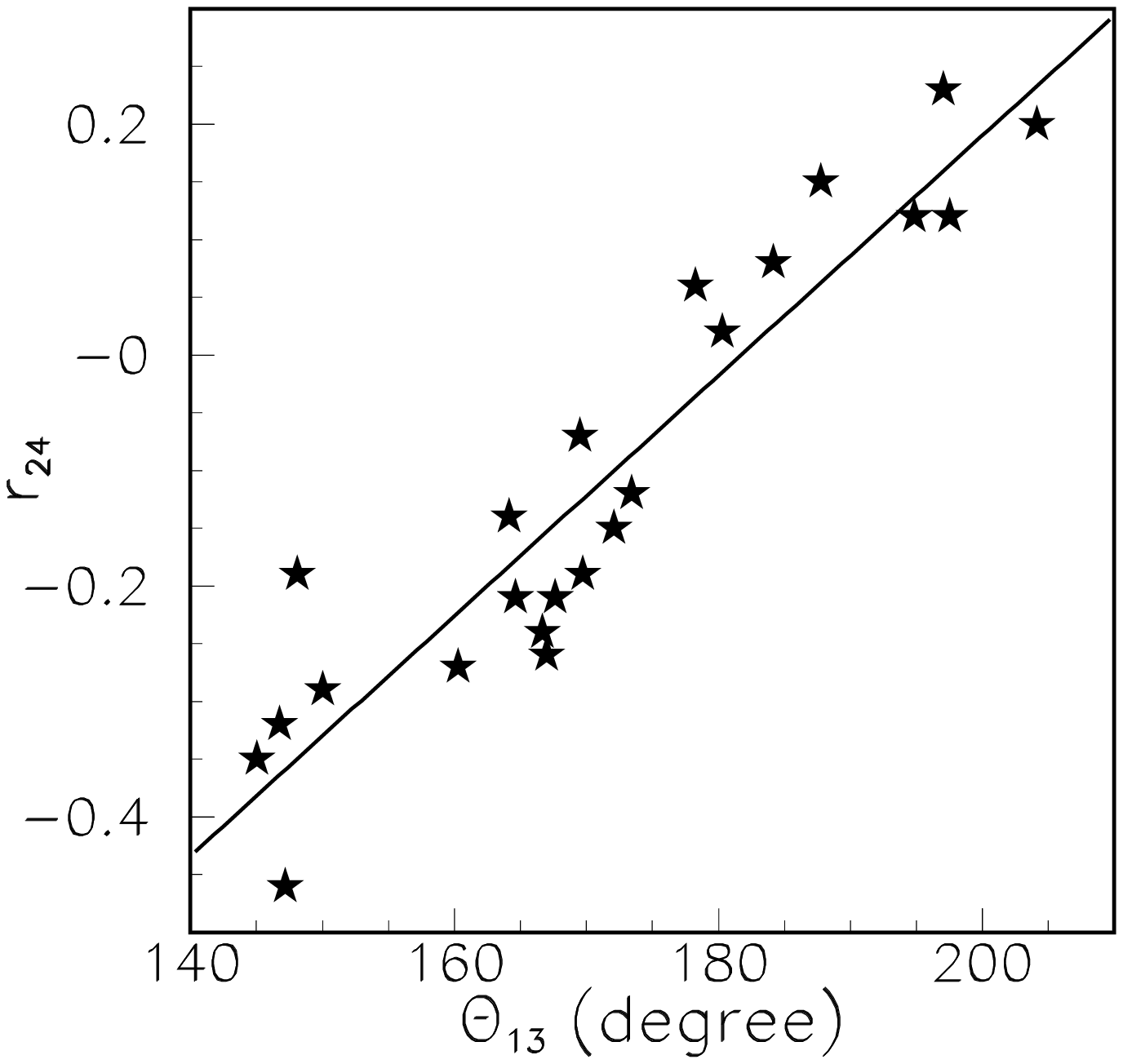}
    \caption{Normalised coordinates of the \citet{castles} quad sample. From left to right: distribution of $r$; correlation between $\theta_{24}$ (abscissa) and $r_{13}$ (ordinate); correlation between $\theta_{13}$ (abscissa) and $r_{24}$ (ordinate). The lines show the relations illustrated in Table \ref{tab2}.}
  \label{fig2}
\end{figure*}
The  reason  for  using  polar  coordinates  is that they are better adapted to describe correlations such as suggested by \cite{Saha2003}. We note in particular the  presence  of  a  strong  correlation  between  the  opening angle  of  a  pair  of  images  of  equal  parity  (1$-$3  or  2$-$4)  and  the  difference  between  the $r$ values  of  the  other pair.  Figure \ref{fig2}  displays  the  correlation  between $\theta_{13}$=$\theta_3$ and $r_{24}$=$r_{2}-r_4$ together  with  the  correlation  between \mbox{$\theta_{24}$=360$\text{\dego}$+$\theta_2-\theta_4$} and $r_{13}$=$r_1-r_3$.  Other combinations of normalised coordinates have been considered and found to display significant correlation, however much weaker. Part of the observed correlation is trivial and simply results from the definition of normalised coordinates.  A qualitative illustration is shown in Figure 3, which displays the correlation plots of Figure 2 for a random distribution of images in a region of the sky covering a solid angle of similar size as that covered by the \citet{castles} quad sample. While significantly weaker, the correlation is important. A quantitative evaluation of the correlation caused by the lensing mechanism proper is given in Section \ref{sec3}. 
\begin{figure*}
  \centering
    \includegraphics[width=0.3\textwidth,trim=0.cm 1.2cm 0.cm 2.cm,clip]{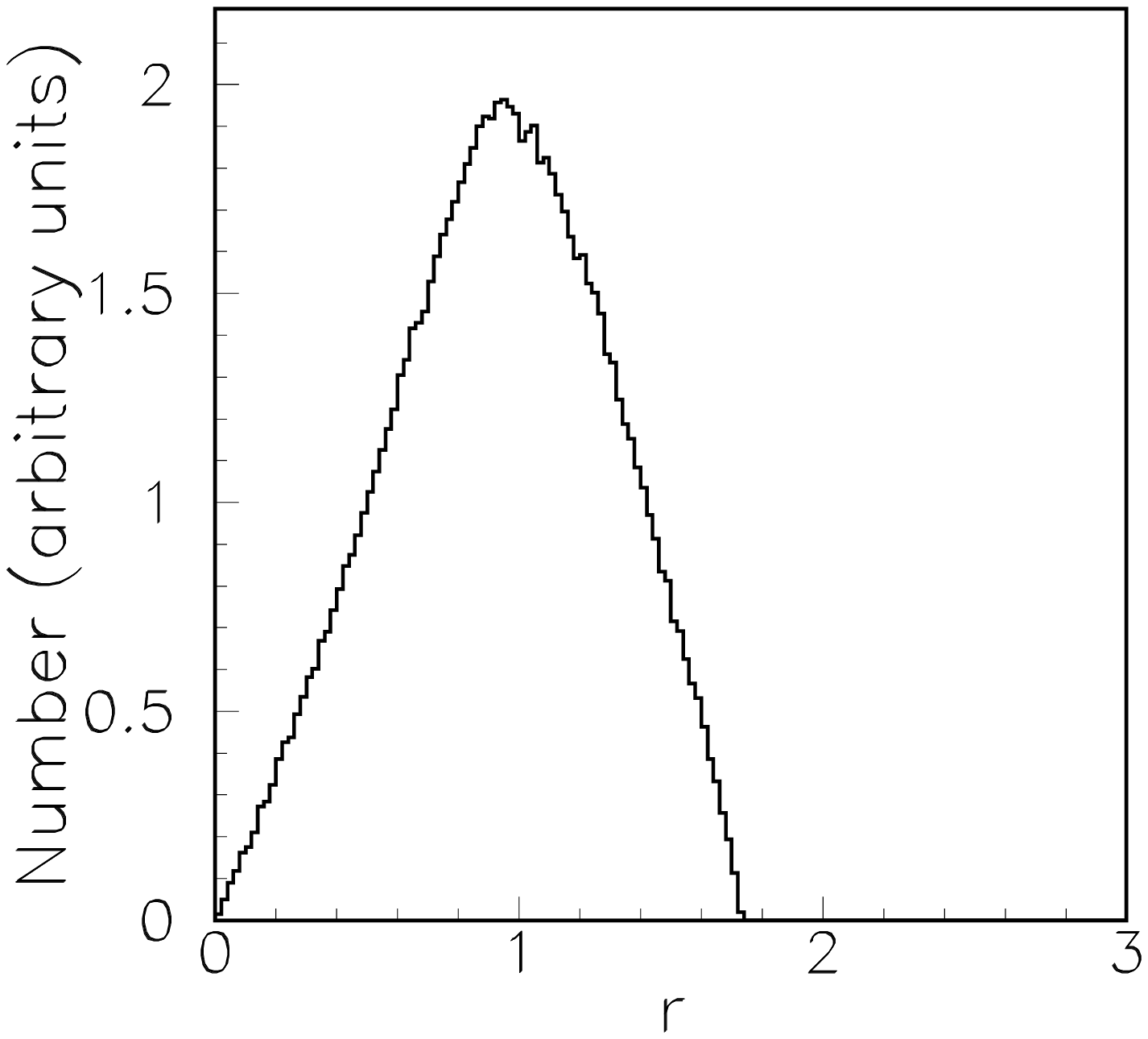}
    \includegraphics[width=0.3\textwidth,trim=0.cm 1.2cm 0.cm 2.cm,clip]{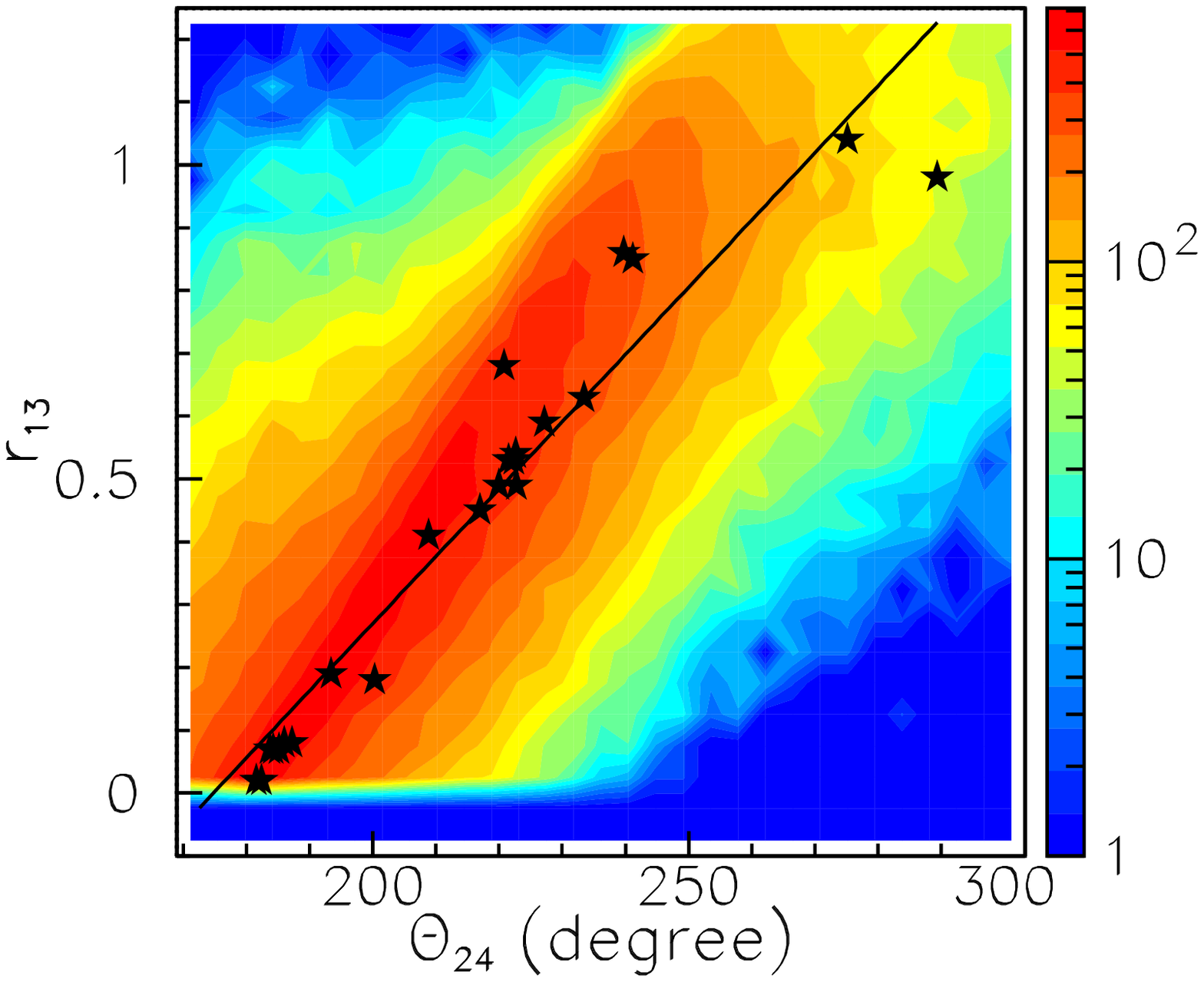}
    \includegraphics[width=0.3\textwidth,trim=0.cm 1.2cm 0.cm 2.cm,clip]{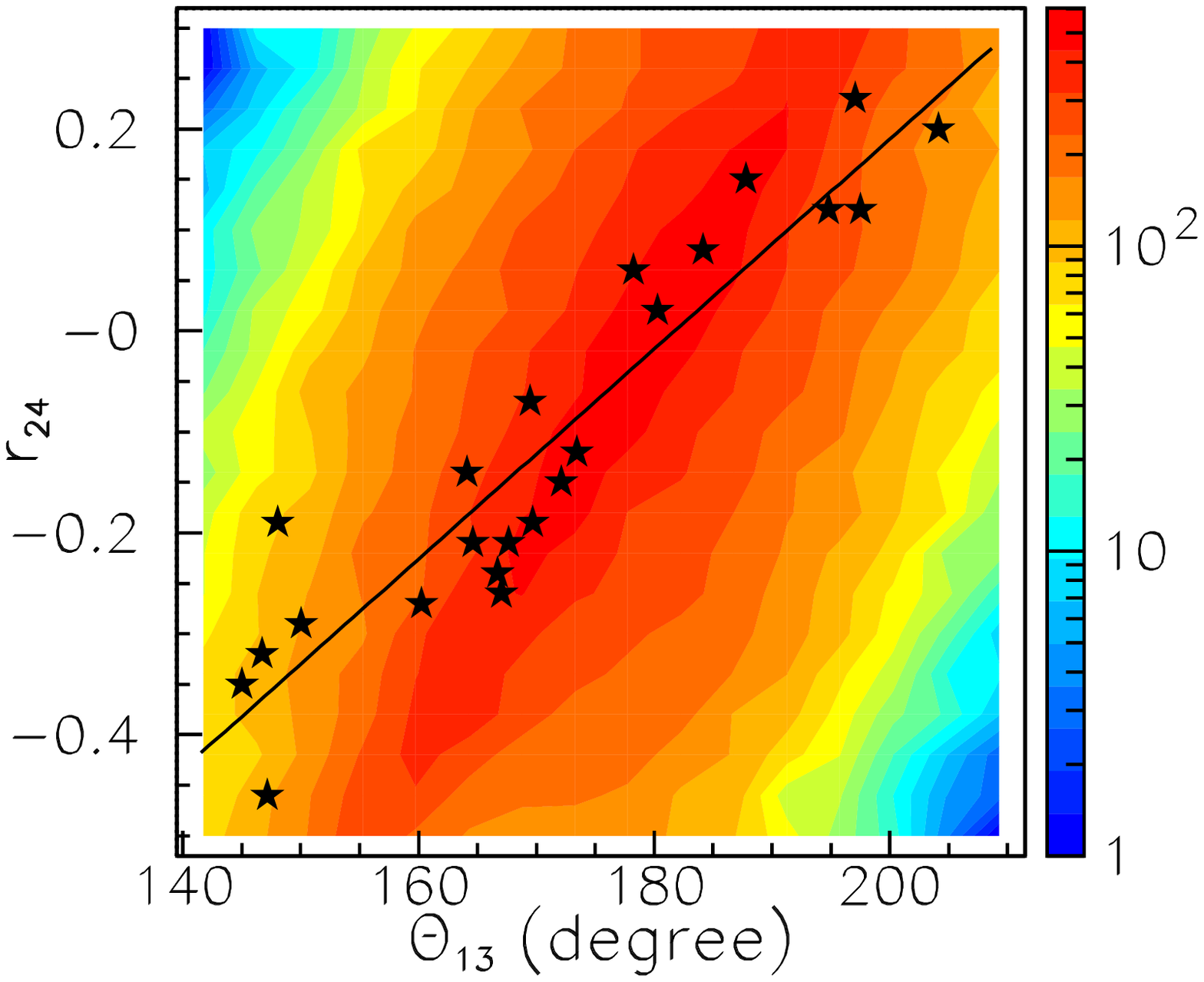}
\caption{Normalised  coordinate  distributions  of  images  uniformly  distributed  in  a  square covering a solid angle of similar size as that covered by the \citet{castles} quad sample.  From  left  to  right: distribution  of $r$;  correlation  between $\theta_{24}$ (abscissa)  and $r_{13}$ (ordinate);  correlation  between $\theta_{13}$ (abscissa)  and $r_{24}$ (ordinate). The \citet{castles} quad sample is overplotted in the two right-most panels.}
\label{fig3}
\end{figure*}
\subsection{Two parameters to define a quad configuration}\label{sec2.3}
The strong correlations illustrated in Figure \ref{fig2} are an invitation to using only two parameters to define the intrinsic configuration of each quad. For example, we might choose $\theta_{13}$ and $\theta_{24}$ and obtain approximate values of $ r_{13}$ and $r_{24}$ by assuming that the observed correlations are exactly obeyed. Writing the correlations in  the  form $r_{ij}=a_{ij}\theta_{kl}+b_{ij}$ the  values  of  the  coefficients  are  listed  in  Table \ref{tab2}  together  with  the  values  of \mbox{$\Delta r_{ij}=\{\sum_{1,23}(r_{ij}-a_{ij}\theta_{kl}-b_{ij})^2/23\}^{1/2}$} which measures how well the correlation is obeyed globally by the sample of 23 quads. We also list the equivalent quantity \mbox{$\Delta \theta_{kl}=\{\sum_{1,23}(\theta_{kl}-r_{ij}/a_{ij}+b_{ij}/a_{ij})^2/23\}^{1/2}$}. For each quad, we calculate the values of $\theta_{kl}$ and $r_{ij}$, which we call $\theta^*_{kl}$ and $r^*_{ij}$ respectively, which obey exactly the correlation and are as close as possible from $\theta_{kl}$ and $r_{ij}$ within \mbox{$\Delta \theta_{kl}$ and $\Delta r_{ij}$. Precisely,} 
\begin{align*}
\theta^*_{kl}=&\theta_{kl}+a_{ij}(r_{ij}-a_{ij}\theta_{kl}-b_{ij})/[a^2_{ij}+(\Delta r_{ij}/\Delta\theta_{kl})^2]\\
r^*_{ij} =& r_{ij}-(\Delta r_{ij}/\Delta \theta_{kl})^2(r_{ij}-a_{ij}\theta_{kl}-b_{ij})/\\
&[a^2_{ij}+(\Delta r_{ij}/\Delta \theta_{kl})^2]
\end{align*}
The values of ($r_{ij},r^*_{ij}$) and of ($\theta_{kl},\theta^*_{kl}$) are listed in Table \ref{tab1} together with the values of \mbox{$\delta_{ij}=\{(r_{ij}-r^*_{ij})^2$/$\Delta r^2_{ij}+(\theta_{kl}-\theta^*_{kl})^2$/$\Delta \theta^2_{kl}\}^{1/2}$}, which  measure  how  well  the  correlation  is  obeyed  by  each quad separately. 
\begin{table*}
  \centering
  \caption{Parameters defining the observed correlations. They are defined as $r_{ij}=a_{ij}\theta_{kl}+b_{ij}$ or equivalently $\theta_{kl}=r_{ij}/a_{ij}-b_{ij}/a_{ij}$. $\Delta r_{ij}$ and $\Delta\theta_{kl}$ are defined in Section \ref{sec2.3} and measure how well the correlation is obeyed by the \cite{castles} quad sample.}
  \label{tab2}
  \begin{tabular}{c c c c c c c}
    \hline\hline
    $ijkl$
    &$a$
    &$b$
    &$1/a$
    &$-b/a$
    &$\Delta r_{ij}$
    &$\Delta \theta_{kl} (\text{\dego})$\\
   \hline
    1324
    &0.0107
    &$-$1.87
    &94
    &175
    &0.086
    &8.0\\
   
    2413
    &0.0104
    &$-$1.89
    &96
    &182
    &0.069
    &6.6\\
    \hline\hline
  \end{tabular}
\end{table*}
We  note  that  if  the  \citet{castles}  sample  were  large  enough,  symmetry  would  impose  that $\theta_{ij}$=180\text{\dego}  when $r_{kl}$=0, namely $-b_{kl}/a_{kl}$=180\dego.
\begin{figure*}
  \centering
    \includegraphics[width=0.28\textwidth,trim=0.cm 0.cm 0.cm -1.cm,clip]{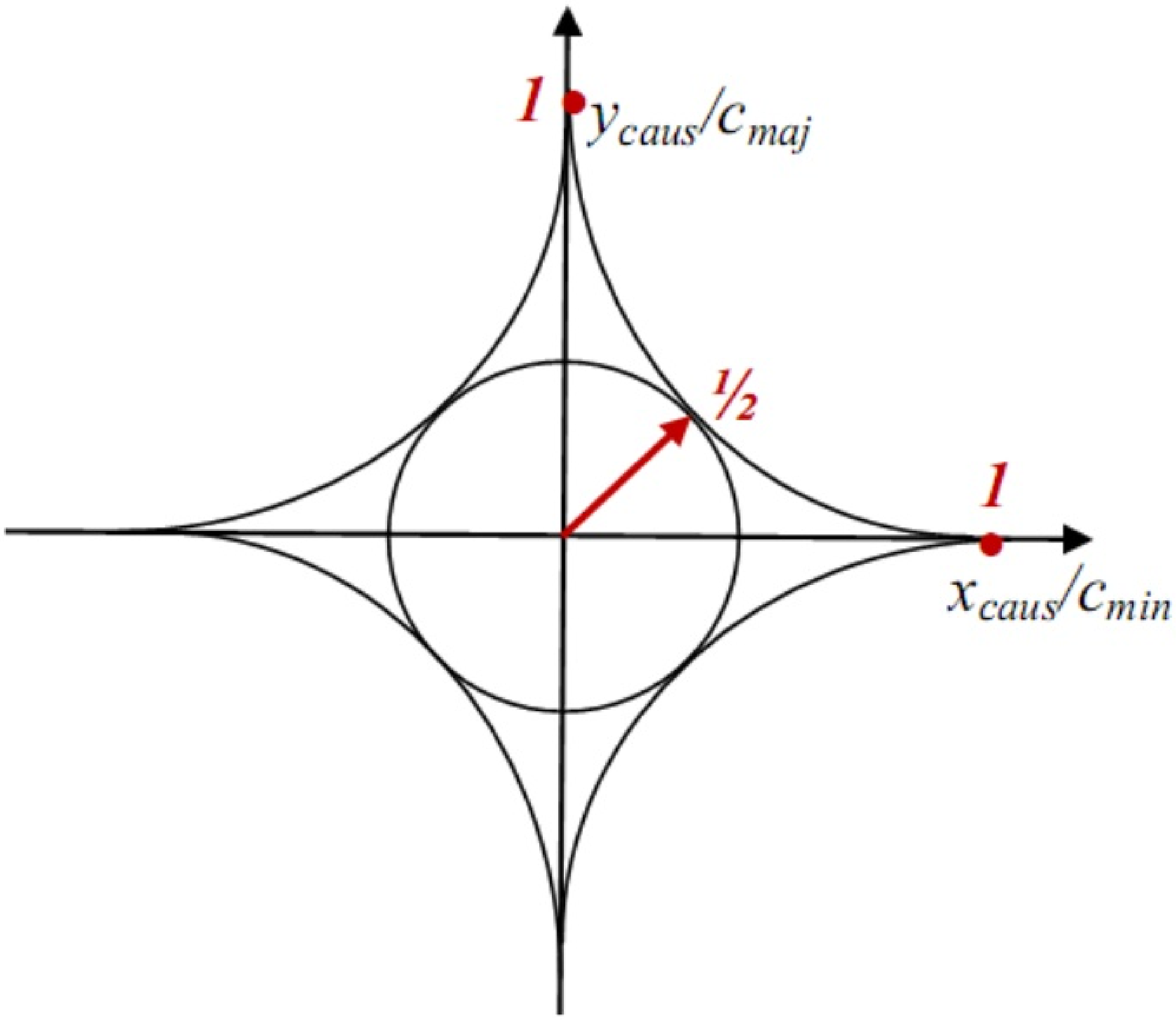}
    \includegraphics[width=0.3\textwidth,trim=0.cm 1.2cm -0.5cm 2.cm,clip]{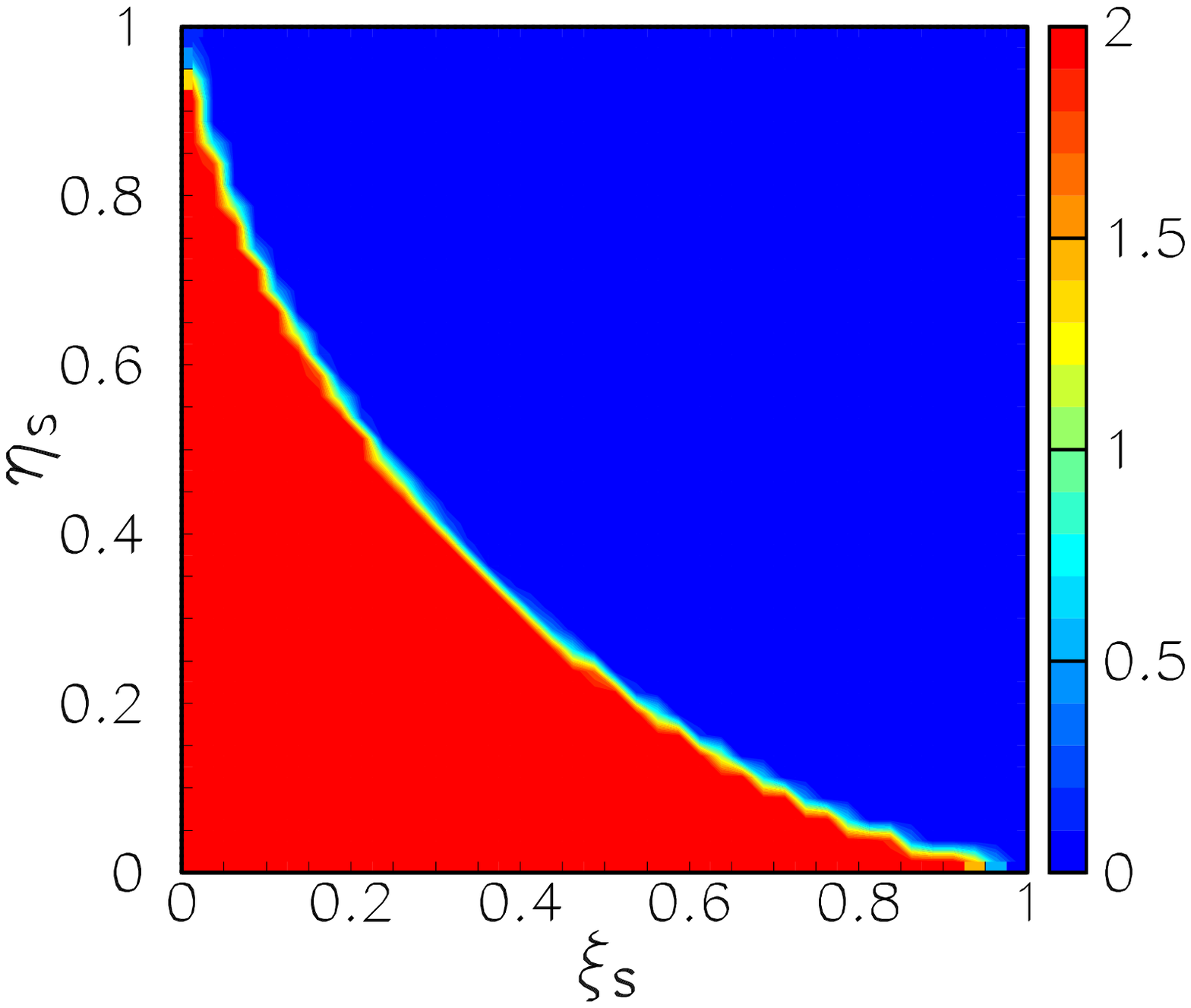}
    \includegraphics[width=0.3\textwidth,trim=0.cm 1.2cm -0.5cm 2.cm,clip]{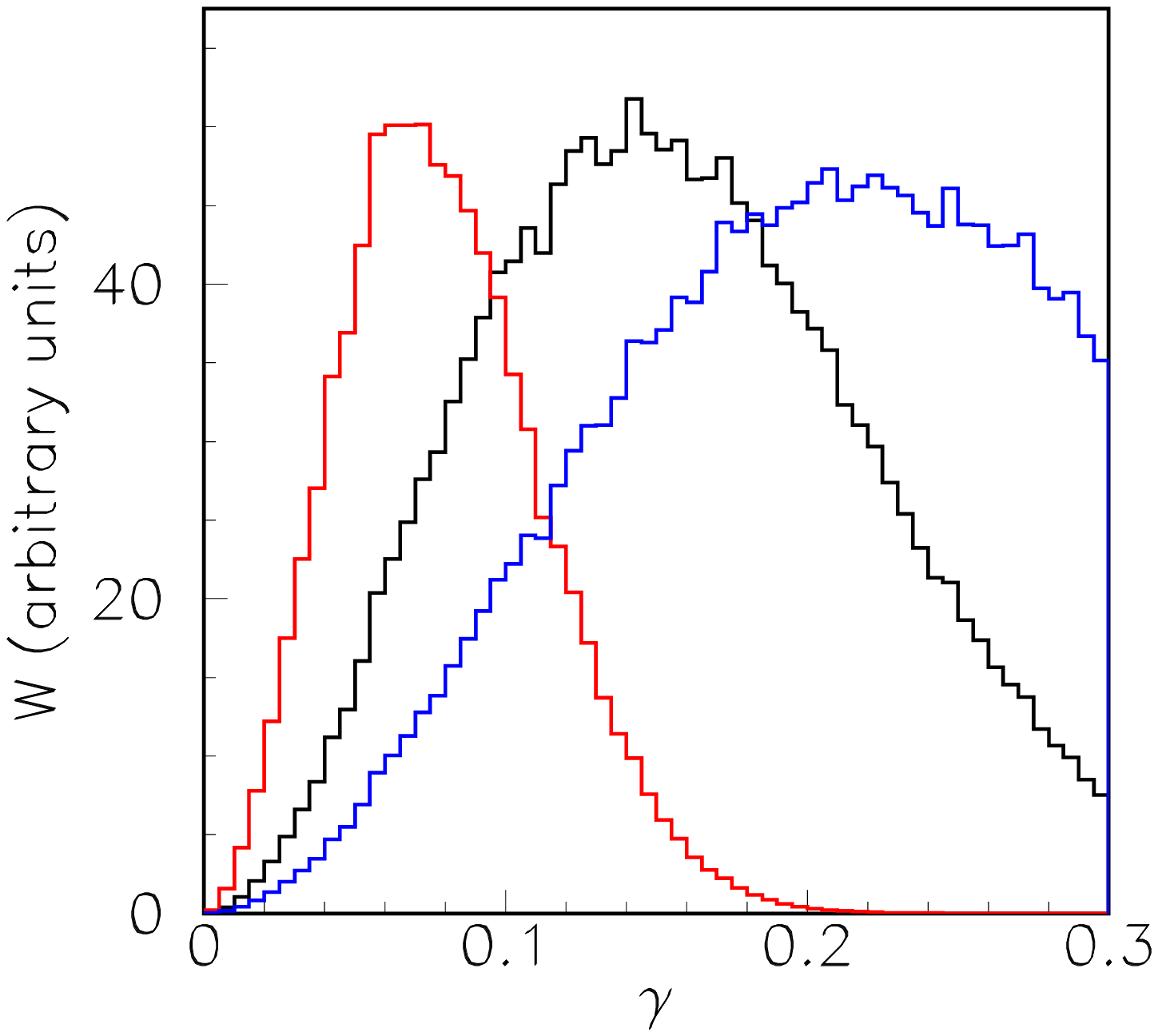}
\caption{Left: the caustic curve in the $y_{caus}$/$c_{maj}$ vs $x_{caus}$/$c_{min}$ plane. Centre: generated distribution in the $\eta_S$ vs $\xi_S$ plane. Right: $\gamma$ distributions for $\sigma_{\gamma}=0.05$ (red), 0.10 (black) and 0.15 (blue). }
\label{fig4}
\end{figure*}
\section{A SIMPLE LENS MODEL}\label{sec3}
\subsection{The model}\label{sec3.1}
We  now  compare  the  prediction  of  a  simple  lens  model  with  the  results  obtained  in  the  preceding section. We assume that the \citet{castles} quad sample consists of quad images of quasar point sources, lensed by a simple potential. The justification for such an assumption is that in most cases the parameters of such a potential  have  effectively  been  obtained  (references  are  given  in the  \citet{castles}  web  site).  The  simplest deviation  from  an  isotropic  potential  of  Einstein radius $R_E$ is  obtained  by  breaking  isotropy,  which  implies two parameters defining respectively the amplitude and position angle of the symmetry breaking term.
This is  usually  done  by  allowing  for  ellipticity  or  by  introducing  an  external  shear.  Here  we  choose  the  latter option and write the lensing potential as \mbox{$\psi=R_E r+1/2 \gamma r^2\cos2(\theta-\theta_{\gamma})$}. It depends on three parameters, the Einstein radius $R_E$ and  the  external  shear $\gamma$ at  position angle $\theta_{\gamma}$.  As $R_E$ fixes  the  scale,  we  can  take  it  equal  to  unity since  we  work  in  normalised  coordinates;  and  for  the  same  reason  the  direction  to  which $\gamma$ is  pointing  is irrelevant: we set $\theta_{\gamma}$=0 and our lens model effectively depends on a single parameter, the \mbox{amplitude $\gamma$ of the external shear.}

The equation of the caustic is given in parameterised form (parameter $\varphi$):
\begin{equation*}
x_{caus}=\{2\gamma/(1+\gamma)\}\cos^3\varphi;\quad y_{caus}=\{2\gamma/(1-\gamma)\}\sin^3\varphi
\end{equation*}
The semi-minor axis of the caustic ($\varphi$=0\dego) is on the $x$ axis with size $c_{min}=2\gamma/(1+\gamma)$ and its semi-major axis  ($\varphi$=90\dego) is  on  the $y$ axis  with  size $c_{maj}=2\gamma/(1-\gamma)$.  Its  equation  reads  then  simply: \mbox{$x_{caus}=c_{min}\cos^3\varphi$}; \mbox{$y_{caus}=c_{maj}\sin^3\varphi$} and \mbox{$\tan\theta=y_{caus}/x_{caus}=(c_{maj}/c_{min})\tan^3\varphi$}. The equation of the critical curve is $r_{cc}=(1-\gamma\cos2\theta_{cc})/(1-\gamma^2)$; its semi-minor and semi-major axes have sizes 1/(1$\pm\gamma$), namely scaled up by a factor (2$\gamma$)$^{-1}$ with respect to those of the caustic. For given values of the source location ($x_S,  y_S$) inside the caustic and of the external shear $\gamma$ we  obtain  the  quad  configuration  in  terms  of  the  normalised  coordinate  parameters  defined  in  the preceding section \mbox{($r_{13}, r_{24}, \theta_{13}$ and $\theta_{24}$)}.

In practice $\gamma$ takes modest values. If it were to require large values it would mean that the lens has a complex structure and the simple form of the potential would probably not be justified. We choose for it a Gaussian distribution of the form exp($-0.5\gamma^2$/$\sigma^2_{\gamma}$) as illustrated in the right panel of Figure \ref{fig4}. As has been amply remarked by previous authors \mbox{\citep{Saha2003, Woldesenbet2012}} the precise value taken by $\gamma$ has little impact on the image configuration once expressed in a form independent of orientation and scale. In what follows we take as default value $\sigma_{\gamma}=0.1$, covering a broad scale of $\gamma$ values \mbox{between 0 and 0.3}.

The size of the quads is at the scale of unity, as is the size of the critical curve which the four images bracket. Indeed, averaging  over $\gamma$, $x_S$ and $y_S$,  we  find  that $r$ has  mean$\pm$rms  values  of \mbox{0.97$\pm$0.23}.  When $\gamma$ approaches  0,  the caustic collapses into a single point at the origin and the critical curve becomes the circular Einstein ring on which  the  positions  of  the  four  images  are  no  longer  defined.  In  practice,  for  an  extended  source,  the extension  of  the  images  becomes  such  that  they  cover  the  whole  ring.  The  probability  for  a  quasar  to  be located  inside  the  caustic  of a  lens  of  external  shear $\gamma$ is  proportional  to  the  area  of  the  caustic,  namely proportional  to $\gamma^2/(1-\gamma^2)$.  Our  sample  of  model  quads  (Figure  \ref{fig4})  is  therefore  obtained  by  generating  a uniform  source  distribution  inside  a  caustic  of  shear $\gamma$,  also  uniformly  distributed, and  giving  each quad  a weight $W=\gamma^2(1-\gamma^2)^{-1}\text{exp}(-0.5\gamma^2/\sigma^2_{\gamma})$.

Rather than $x_S$ and $y_S$, which depend on $\gamma$, we use coordinates that define the position of the source inside the caustic independently from $\gamma$. Precisely, as the quad configuration is independent from the signs of $x_S$ and $y_S$,  we  use  coordinates $\xi_S=|x_S|/c_{min}$ and $\eta_S=|y_S|/c_{maj}$.  Both  take  values  between  0  and  1  and  are uniformly distributed in the plane (Figure \ref{fig4}).

\subsection {Model predictions: general features}\label{sec3.2}
\begin{figure*}
  \centering
    \includegraphics[width=0.33\textwidth,trim=0.cm 1.2cm 0.cm 2.cm,clip]{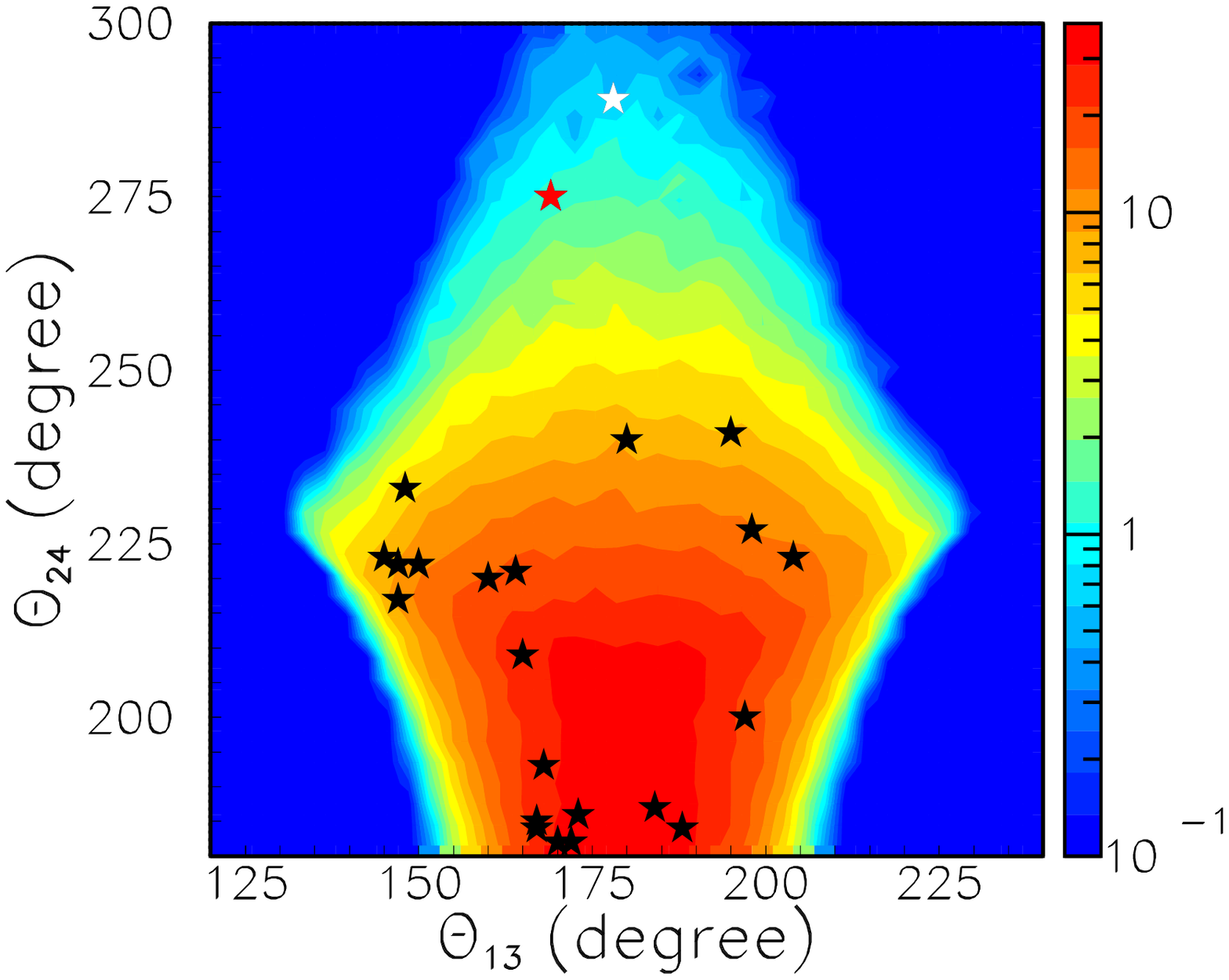}
    \includegraphics[width=0.33\textwidth,trim=0.cm 1.2cm 0.cm 2.cm,clip]{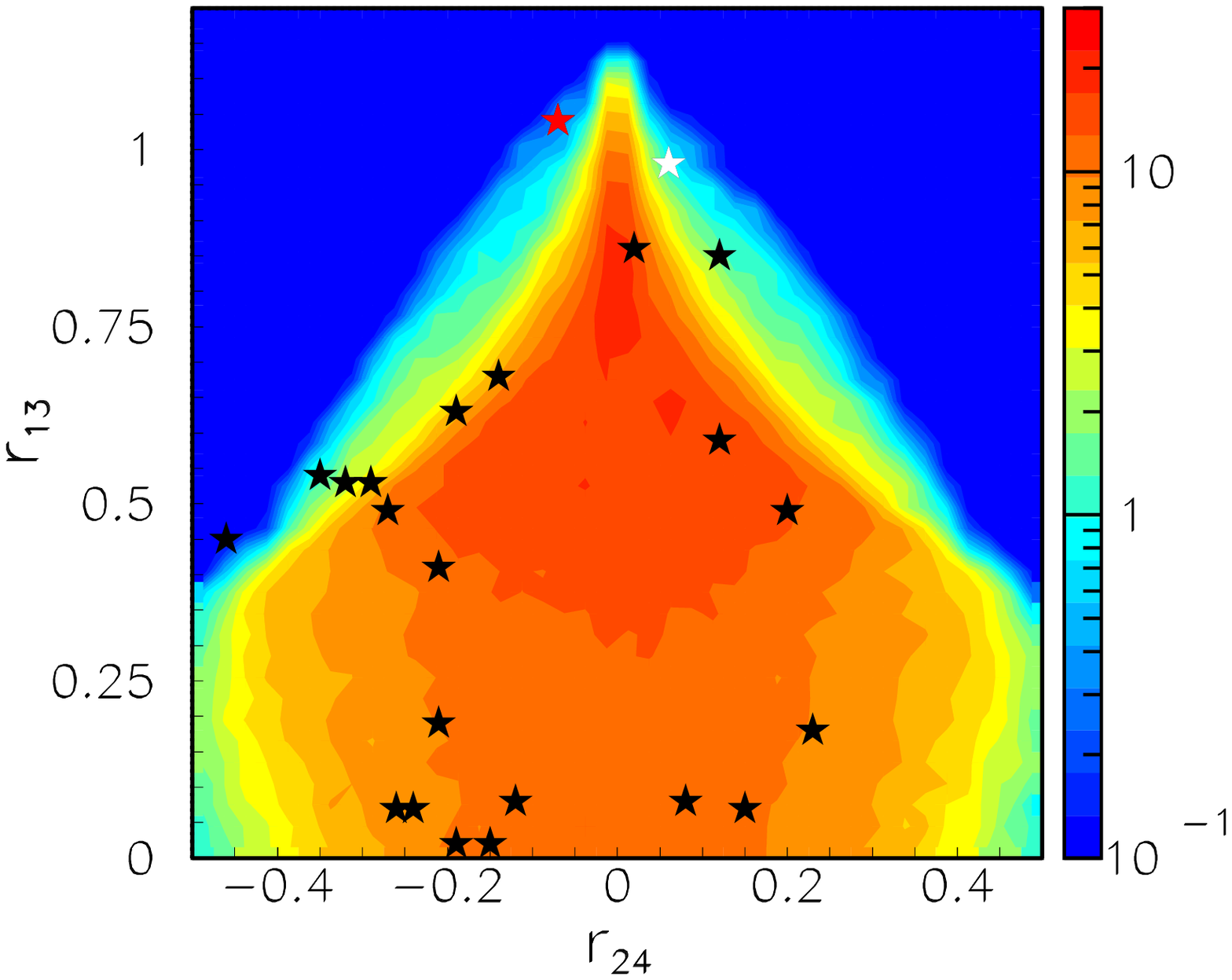}
    \includegraphics[width=0.33\textwidth,trim=0.cm 1.2cm 0.cm 2.cm,clip]{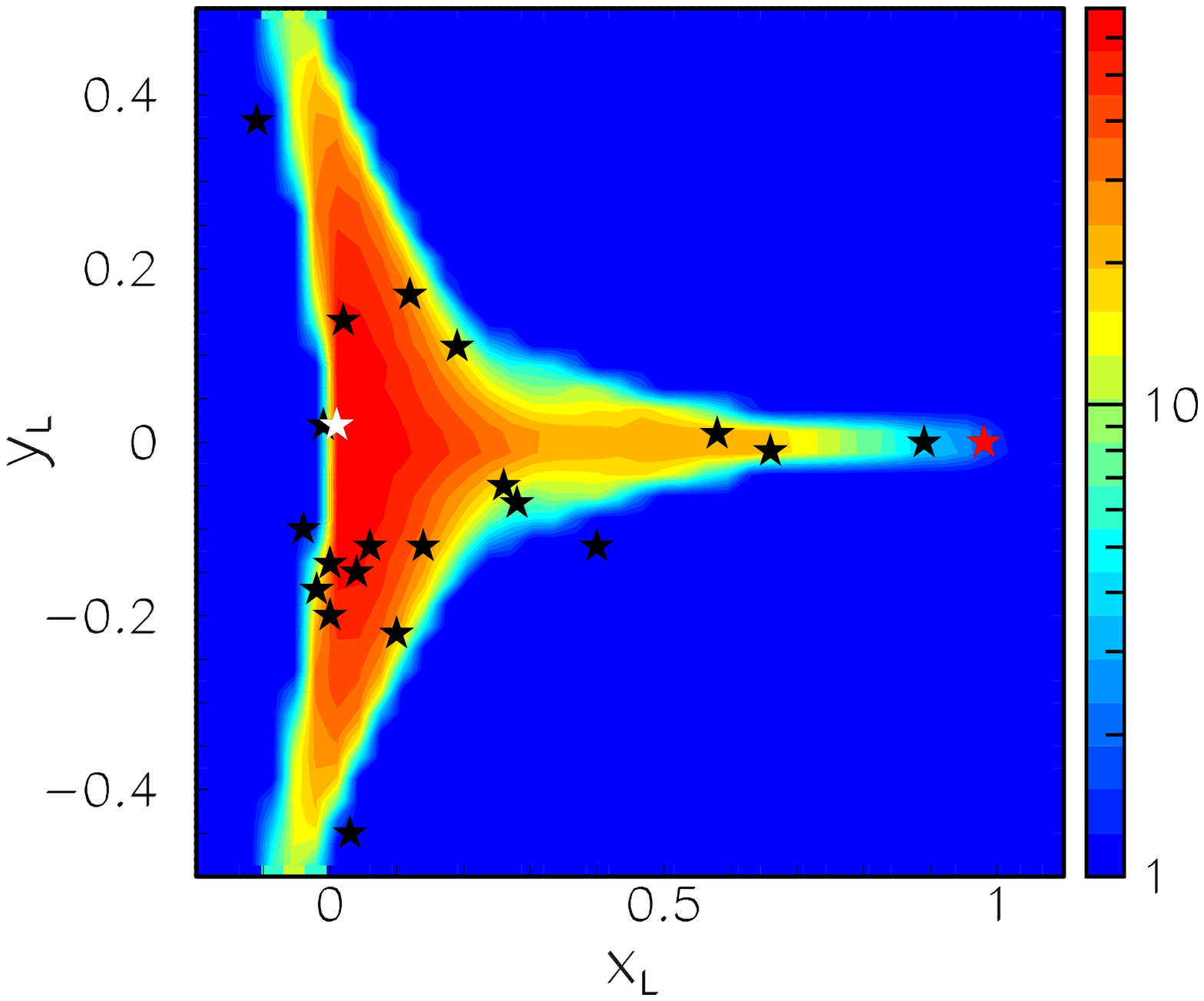}
\caption{Lens model distributions ($\sigma_{\gamma}$=0.1) in the $\theta_{24}$ vs $\theta_{13}$ plane (left), in the $r_{13}$ vs $r_{24}$ plane (centre) and in the $y_L$ vs $x_L$ plane (right, see Section \ref{sec3.4}). The stars show the \citet{castles} quads (white and red stars are for nr 7 and 22 respectively, see Section \ref{sec3.3}). Units are arbitrary.}
\label{fig5}
\end{figure*}

Figure \ref{fig5}   compares  the  quad  configurations  predicted  by  the  model  for $\sigma_{\gamma}$=0.1  with  those of  the \citet{castles} sample.  Axial symmetry is observed as expected about $\theta_{13}$=180\dego, $r_{24}$=0 and $y_{L}$=0.

\begin{figure*}
  \centering
    \includegraphics[width=0.33\textwidth,trim=0.cm 1.2cm 0.cm 2.cm,clip]{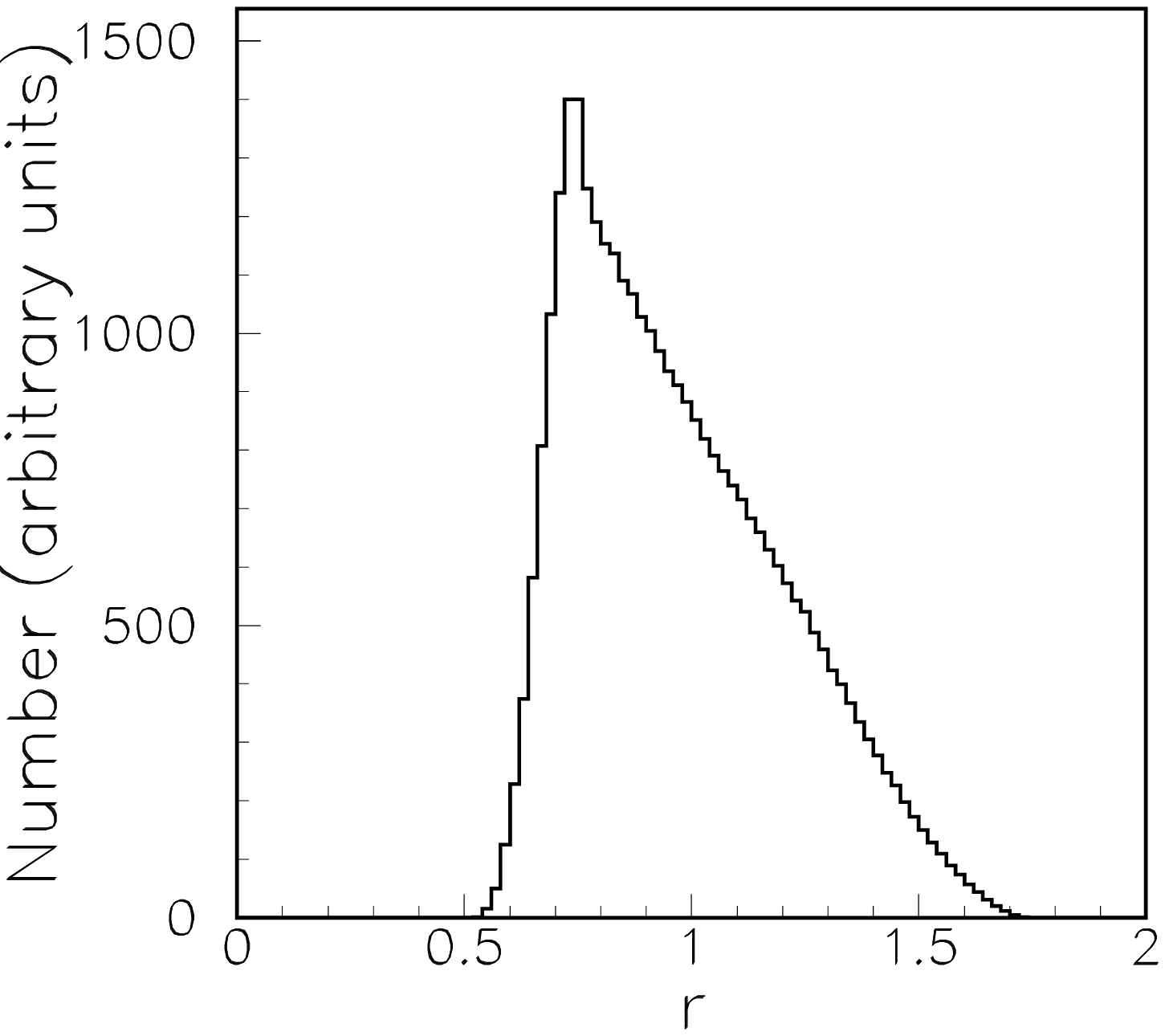}
    \includegraphics[width=0.33\textwidth,trim=0.cm 1.2cm 0.cm 2.cm,clip]{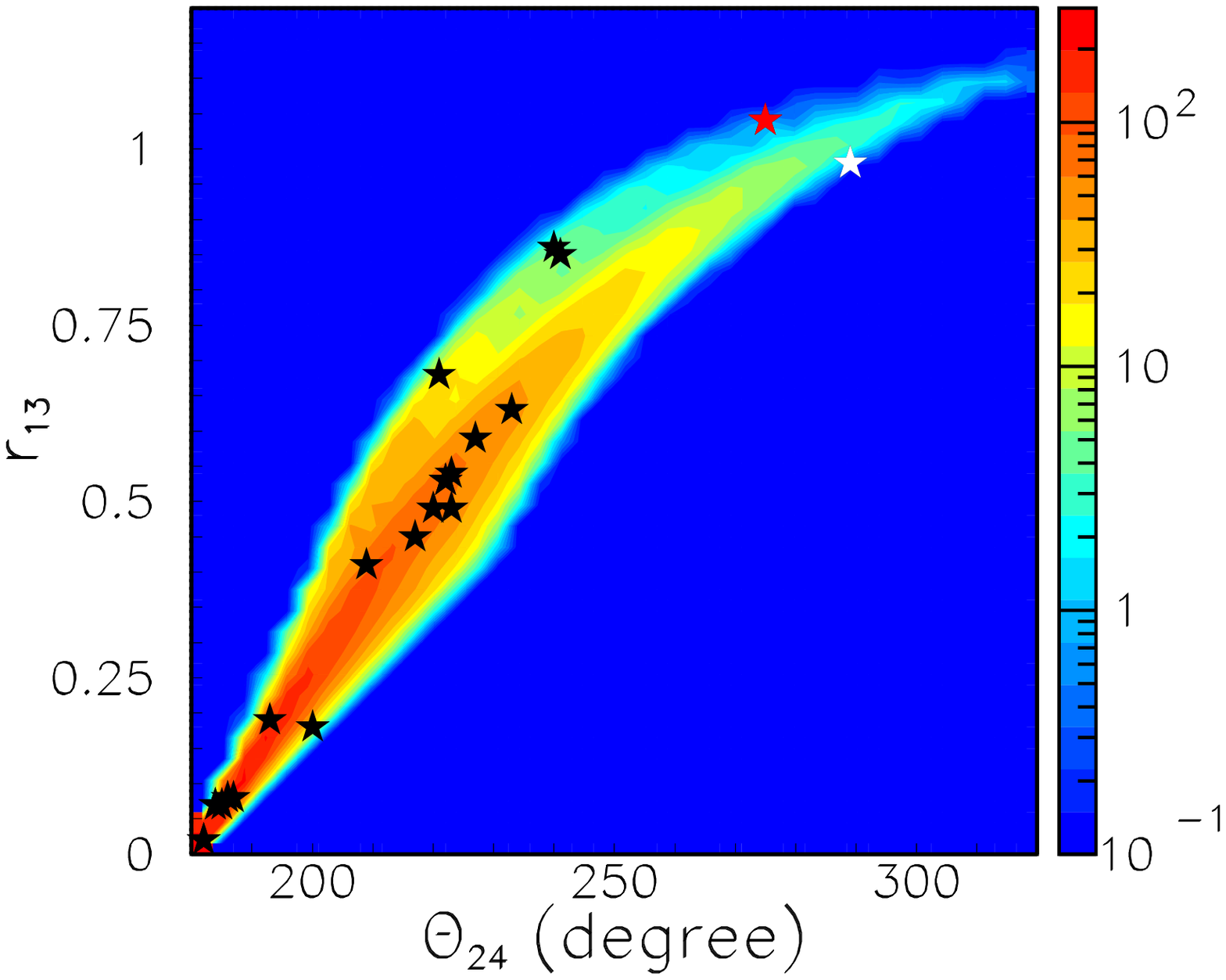}
    \includegraphics[width=0.33\textwidth,trim=0.cm 1.2cm 0.cm 2.cm,clip]{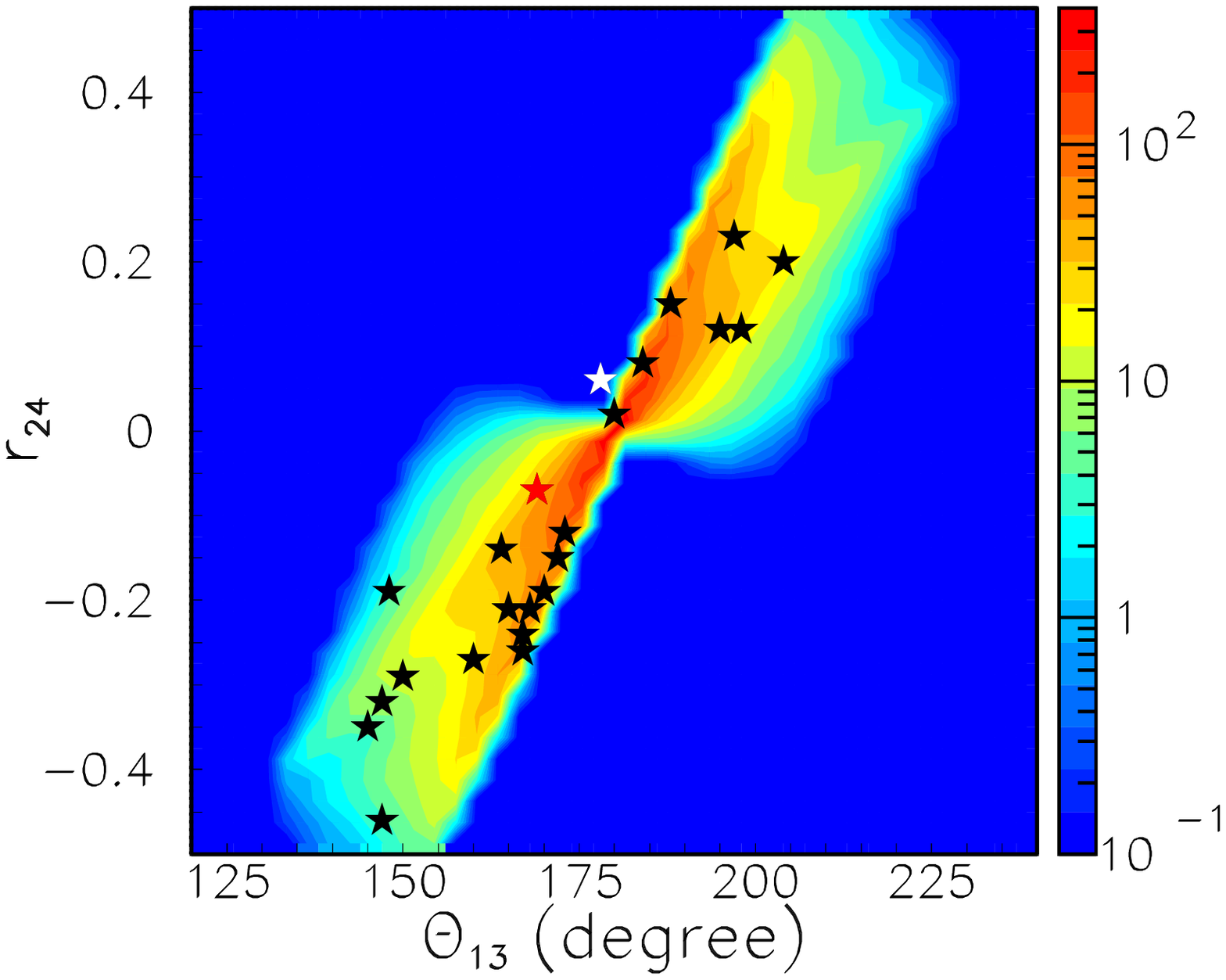}
\caption{Lens  model  correlations. From  left  to  right:  distribution  of $r$;  correlation  between $\theta_{24}$ (abscissa)  and $r_{13}$ (ordinate); correlation between $\theta_{13}$ (abscissa) and $r_{24}$ (ordinate). The stars show the \citet{castles} quads (white and red stars are for nr 7 and 22 respectively, see Section \ref{sec3.3}). Units are arbitrary.}
\label{fig6}
\end{figure*}

Figure  \ref{fig6}  shows  that  the  normalised  coordinate  parameters  of  the  modelled  images  are  strongly correlated in the same way as observed in the \citet{castles} quad sample. While expected, this result gives confidence in the validity of the interpretation of the \cite{castles} quad sample as quadruply-imaged point sources by a simple lens potential. We have checked that changing $\sigma_{\gamma}$ to 0.05  or  0.15  instead  of  0.1  has  little  influence  on  this  result.
\begin{figure*}
  \centering
    \includegraphics[width=0.26\textwidth,trim=0.cm .9cm 0.cm 0.cm,clip]{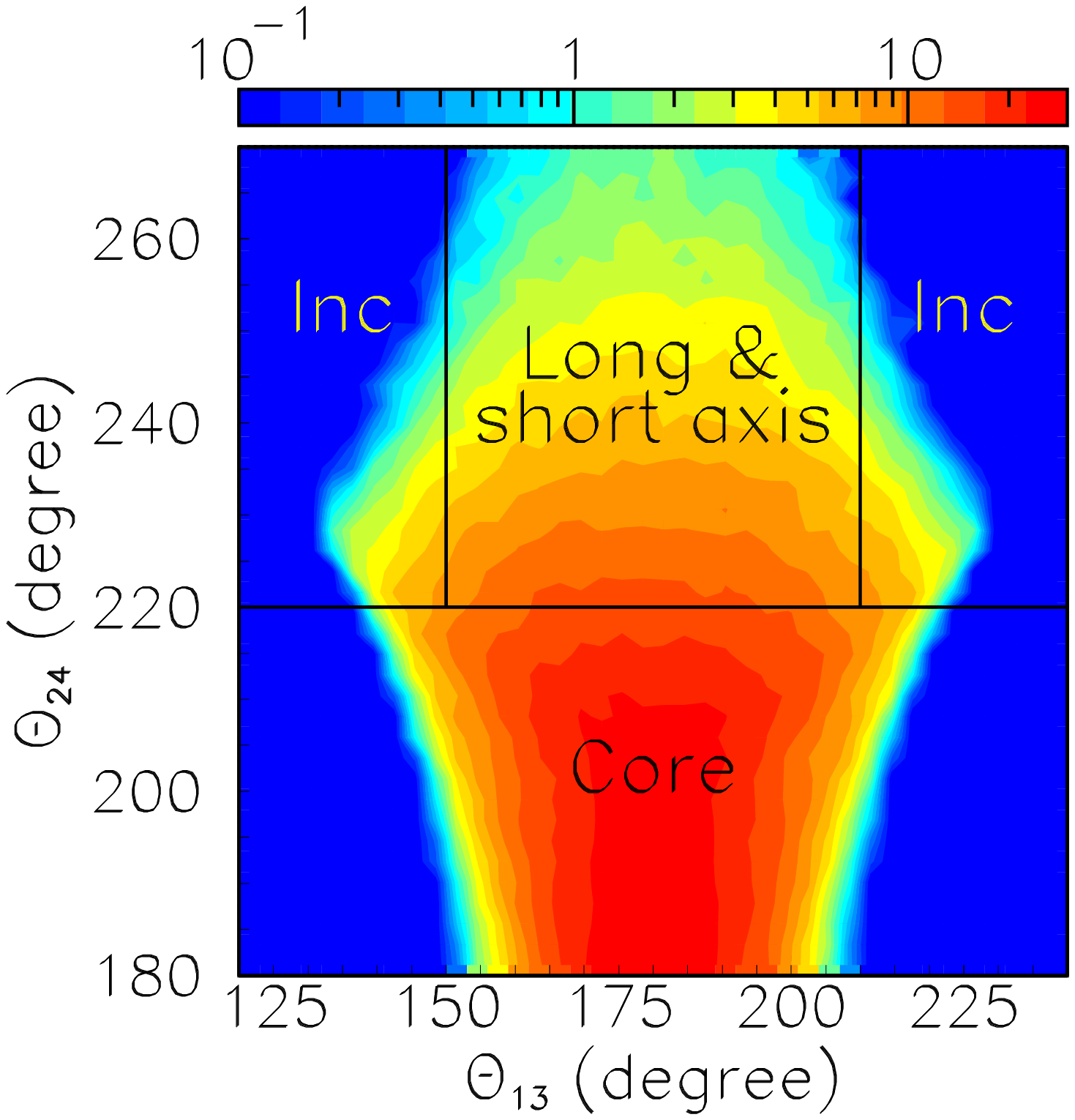}
    \includegraphics[width=0.68\textwidth,trim=0.cm 1.5cm 0.cm 0.cm,clip]{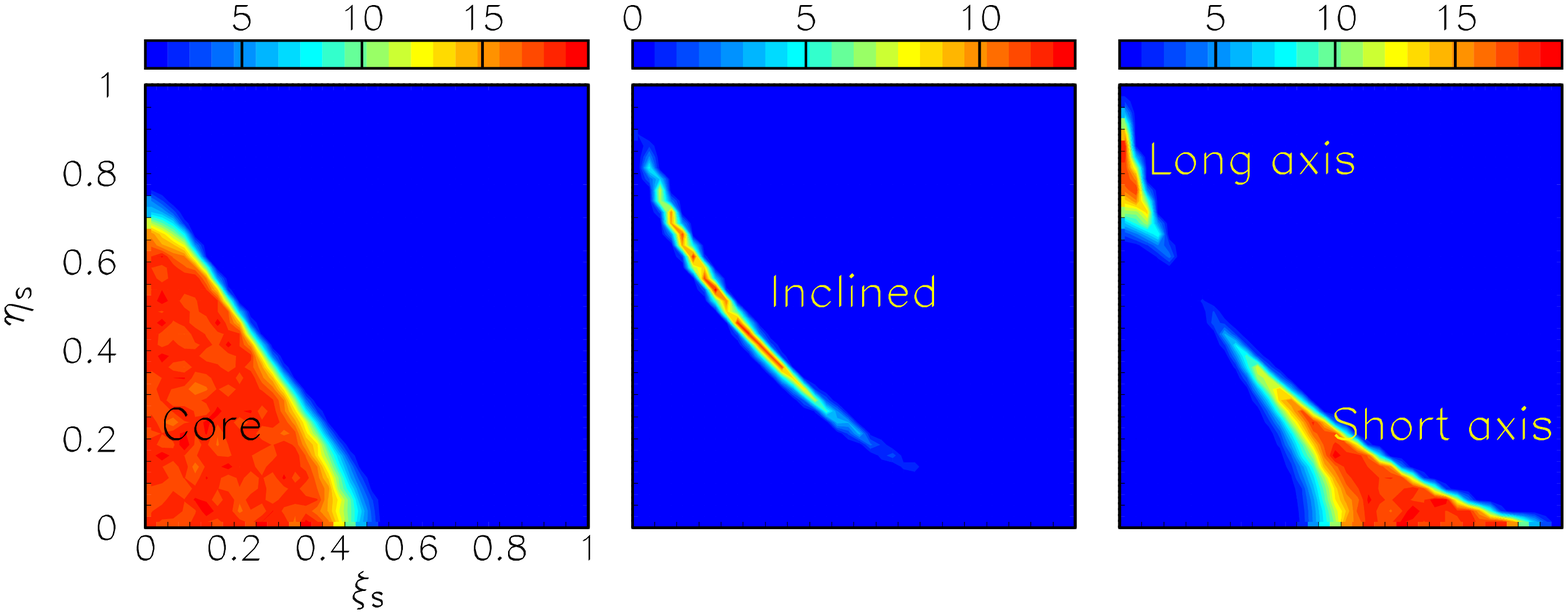}
\caption{Saha \& Williams classification: correlation between the quad configuration in the $\theta_{24}$ vs $\theta_{13}$ plane (left panel) and  the  location  of  the  source  in  the $\eta_S$ vs $\xi_S$ plane. The  three right-most  panels  are, from left to right,  for  core  quads, inclined quads and short or long axis quads respectively, as defined in the left panel: core quads as $\theta_{24}$$<$220\dego, inclined quads as $\theta_{24}$$>$220\dego \& $|\theta_{13}-180\text{\dego}|$$>$30\dego, long and short axis quads as $\theta_{24}$$>$220\dego \&  $|\theta_{13}-180\text{\dego}|$$<$30\dego. The model uses $\sigma_{\gamma}$=0.1.}
\label{fig7}
\end{figure*}
\begin{figure*}
  \centering  
    \includegraphics[width=0.3\textwidth,trim=0.cm 1.2cm 0.cm 2.cm,clip]{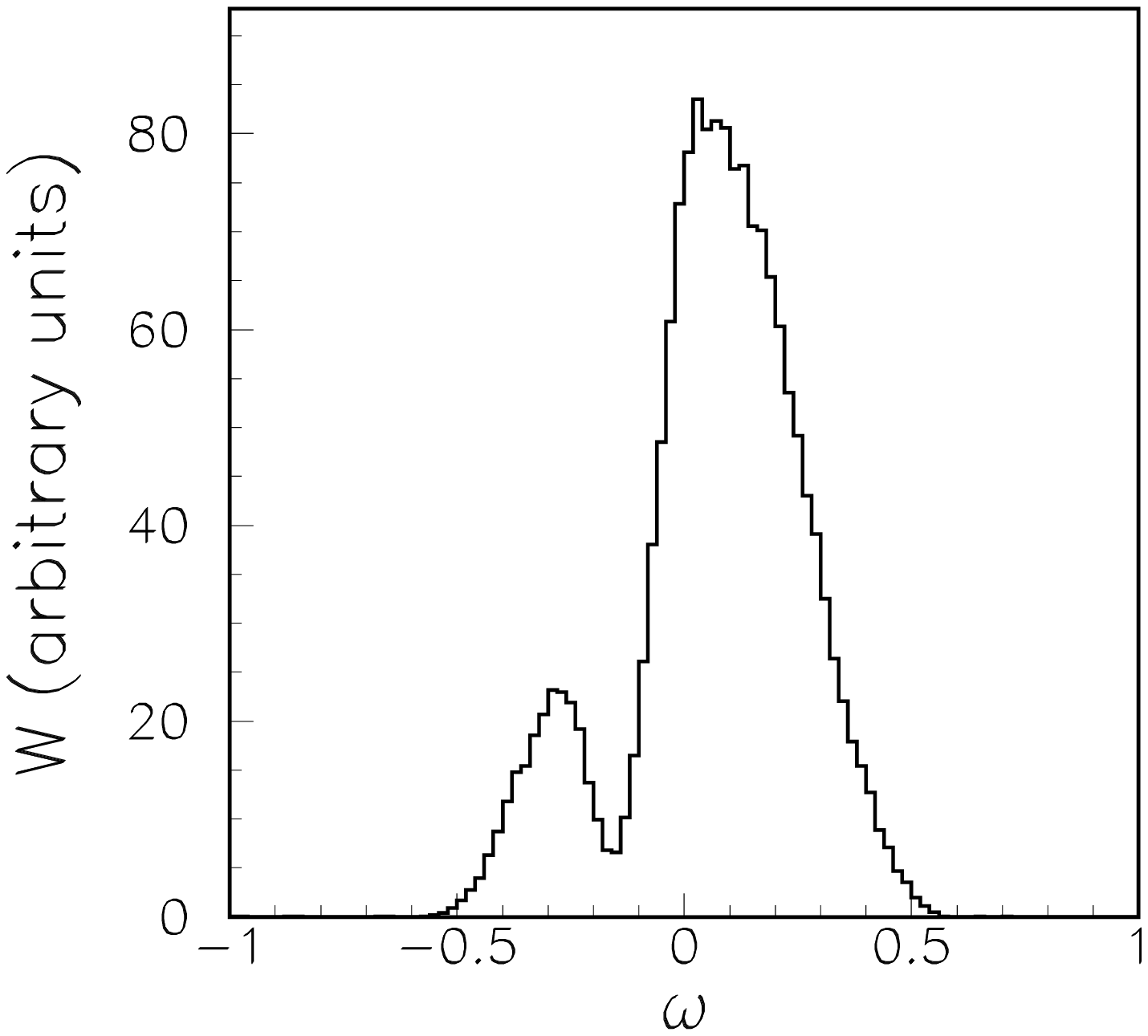}
    \includegraphics[width=0.3\textwidth,trim=0.cm 1.2cm 0.cm 2.cm,clip]{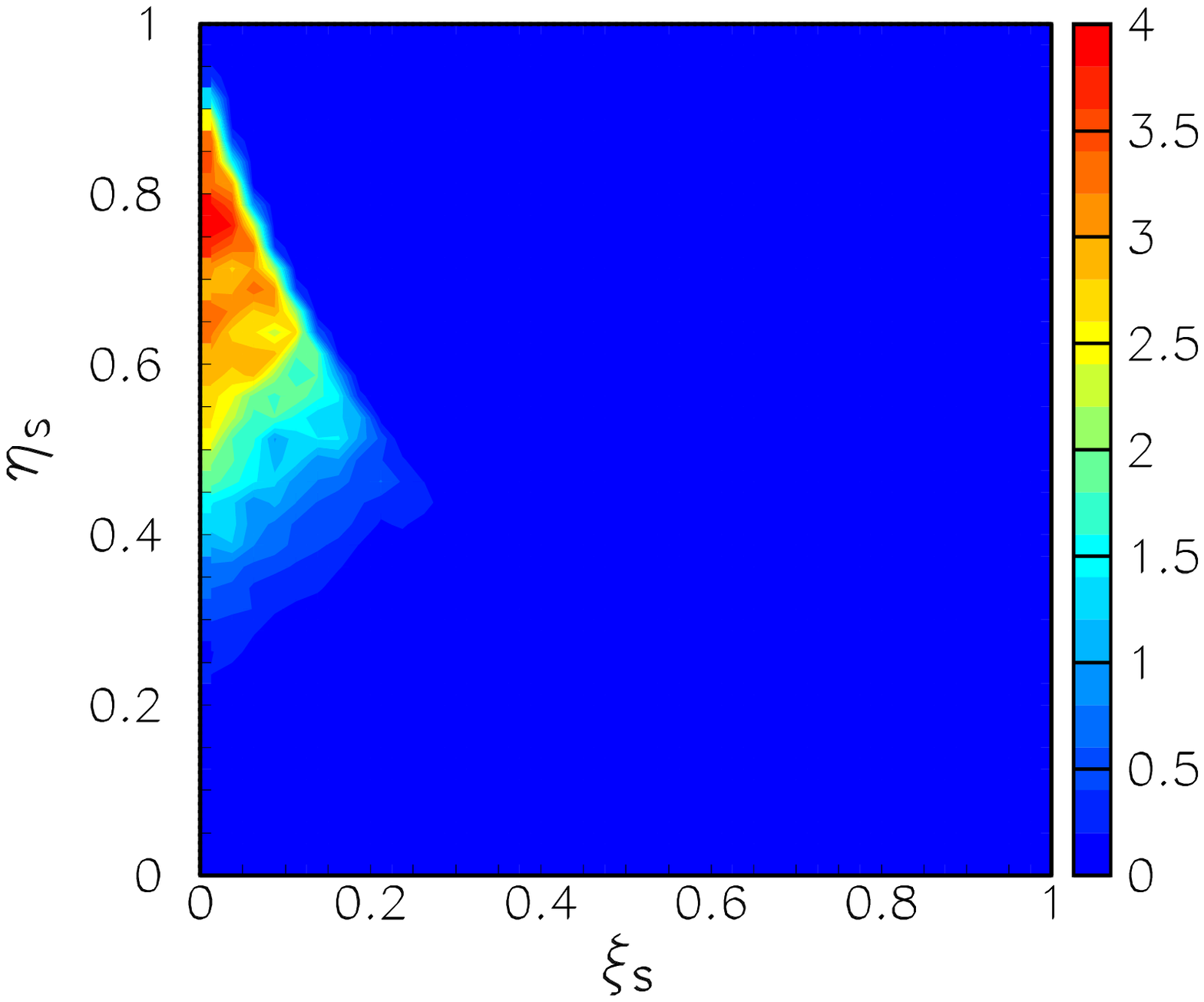}
    \includegraphics[width=0.3\textwidth,trim=0.cm 1.2cm 0.cm 2.cm,clip]{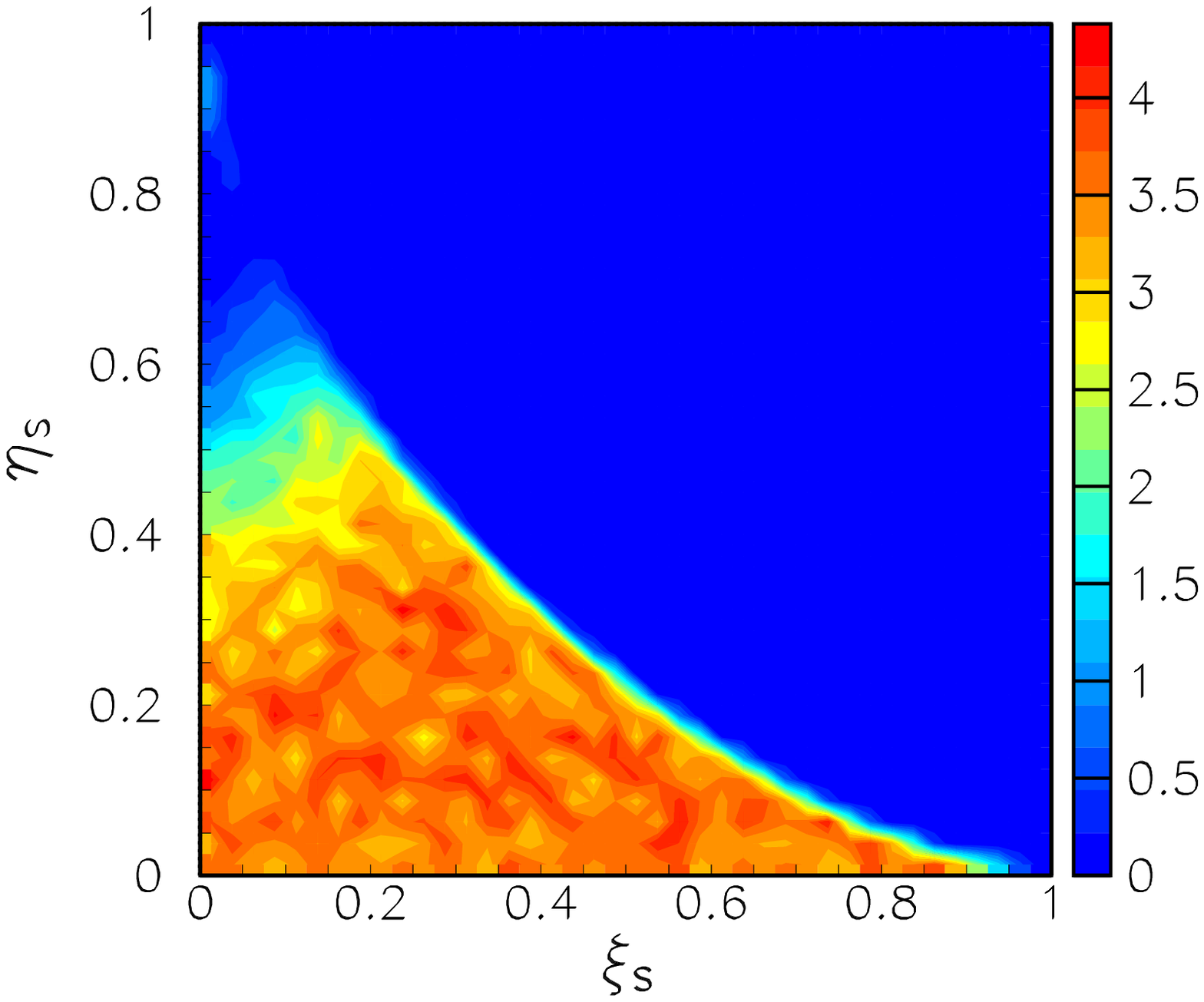}
\caption{Left: distribution of $\omega$. Centre and right: distributions in the the $\eta_S$ vs $\xi_S$ plane of quads having $\omega$$<$$-$0.16 (central panel) or $\omega$$>$$-$0.16 (right panel).}
\label{fig8}
\end{figure*}
The  Saha  \&  Williams classification  implies  a  strong  correlation  between  the  location  of  the  quasar point source inside the lens caustic and the quad configuration. We describe the latter by the four normalised coordinate  parameters, \mbox{$\theta_{24}$, $\theta_{13}$,  $r_{13}$ and $r_{24}$},  approximately  reducible  to  only  two, \mbox{$\theta_{24}$ and $\theta_{13}$} or \mbox{$r_{13}$ and $r_{24}$}. Figure \ref{fig7} illustrates the correlation for quad configurations defined in the $\theta_{24}$ vs $\theta_{13}$ plane. We choose as limits between the core, inclined and cusp classes $\theta_{24}$=220\dego and \mbox{$\theta_{13}$=180\dego $\pm$ 30\dego}; there is of course some arbitrariness in this choice but the precise values are unimportant. Similar results are obtained when defining the quad configuration in the $r_{13}$ vs $r_{24}$ plane, using as limits $r_{13}=0.7$ and $r_{24}=\pm0.05$. There  is  indeed  a  nearly one-to-one correspondence between the three regions defining the quad configuration and the associated regions inside the caustic: the latter do not overlap, except for a small smearing of their edges due to the scattered values taken by $\gamma$.
\begin{figure*}
  \centering
    \includegraphics[width=0.24\textwidth,trim=0.cm 1.2cm 0.cm 2.cm,clip]{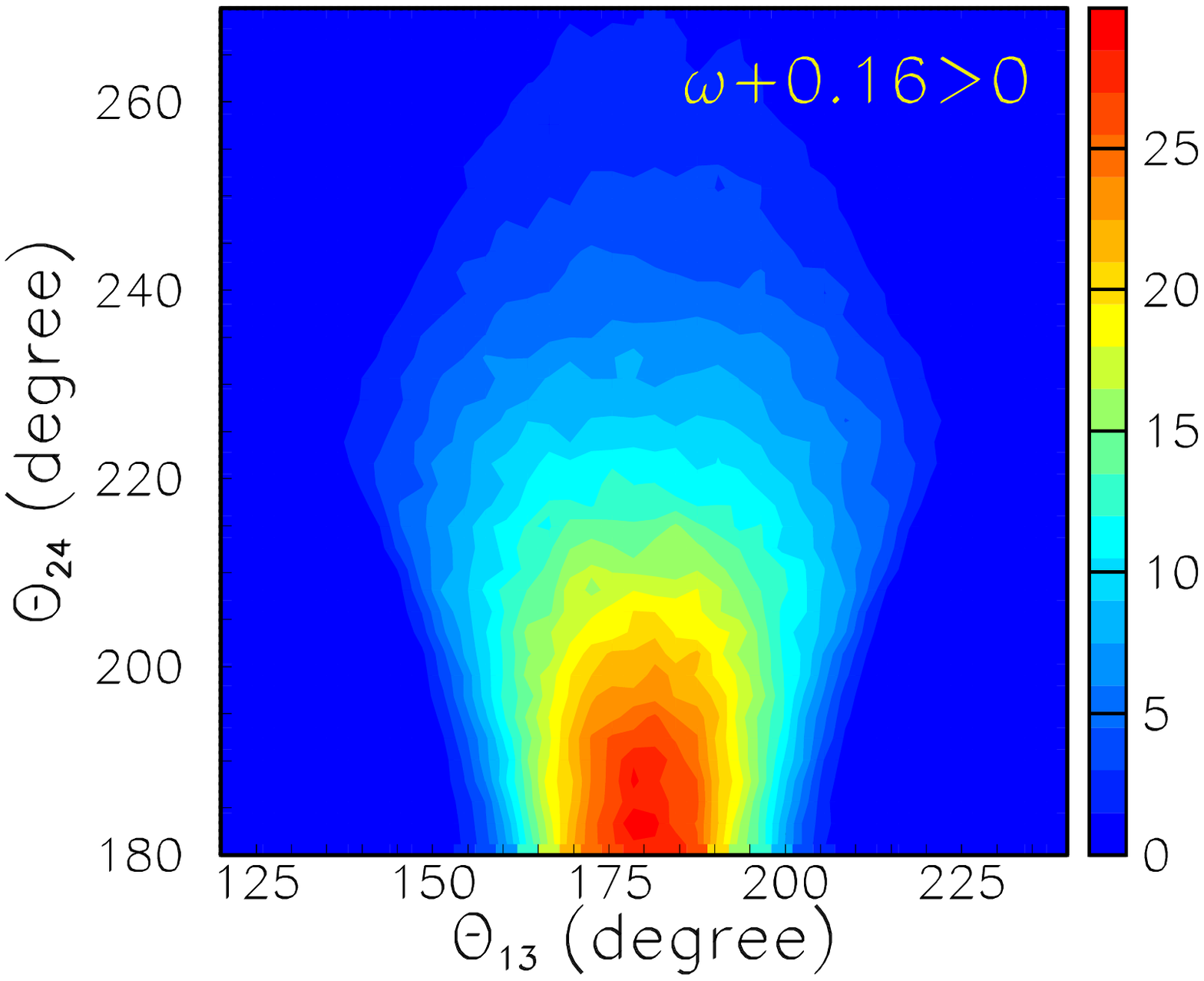}
    \includegraphics[width=0.24\textwidth,trim=0.cm 1.2cm 0.cm 2.cm,clip]{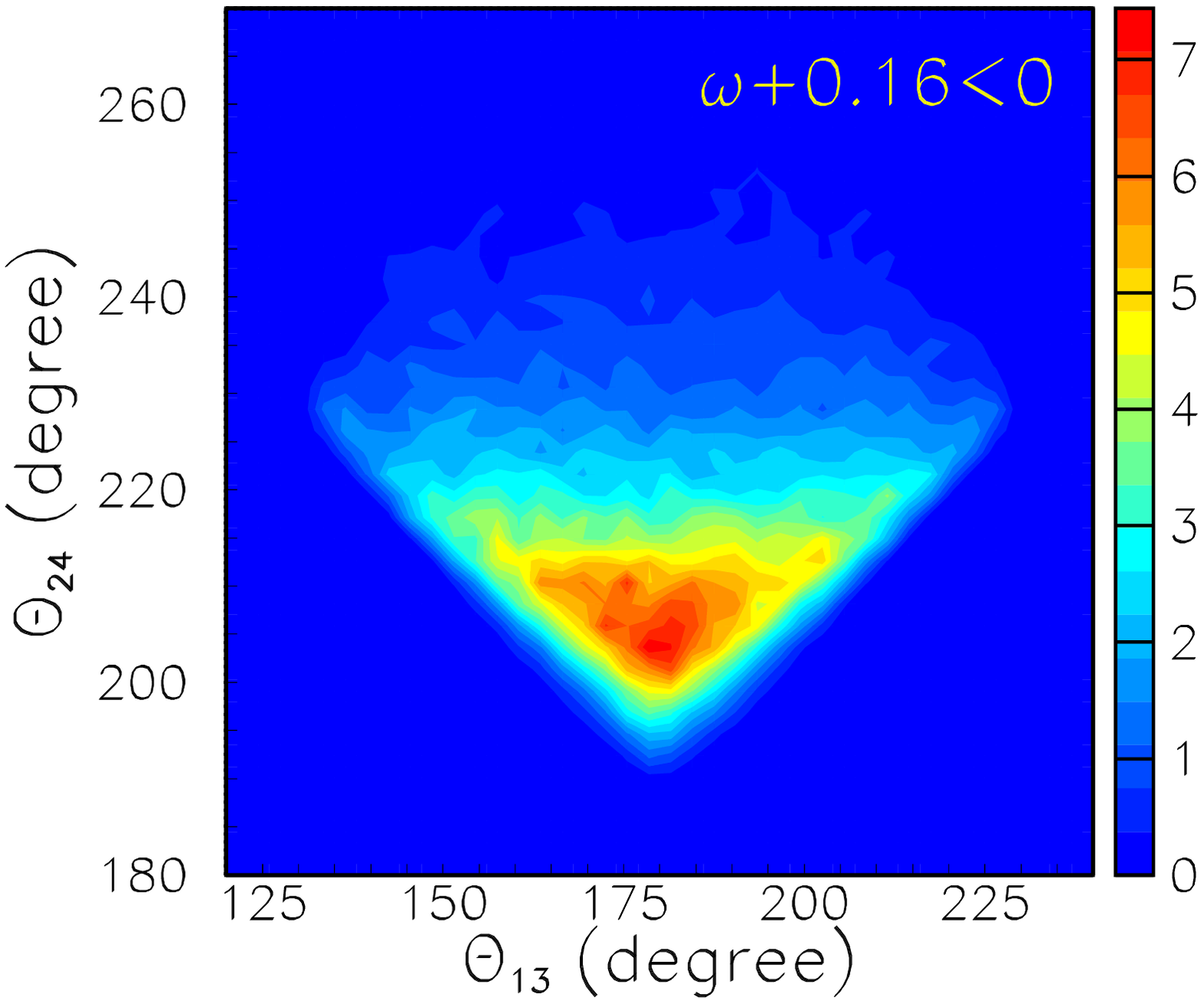}
    \includegraphics[width=0.24\textwidth,trim=0.cm 1.2cm 0.cm 2.cm,clip]{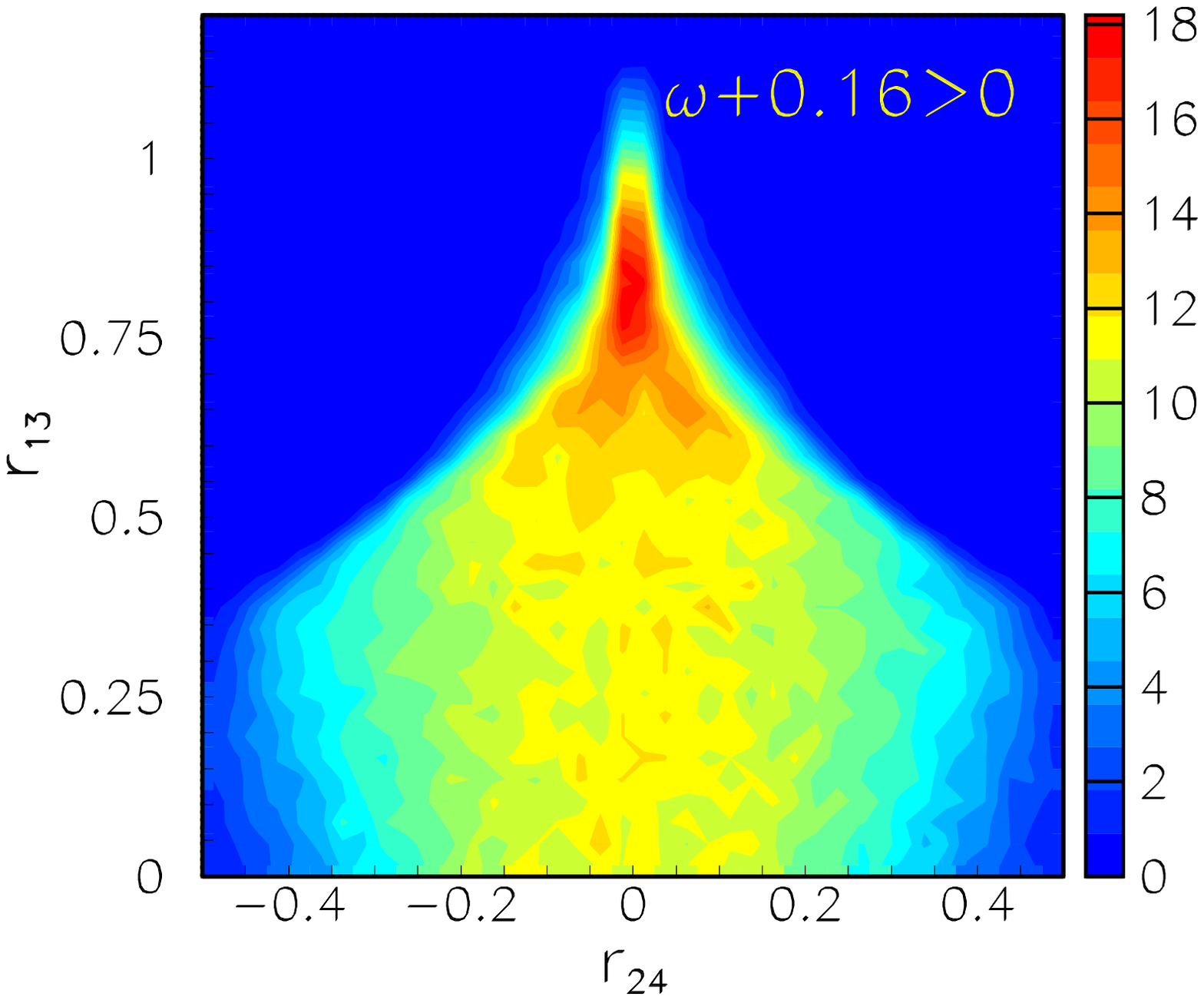}
    \includegraphics[width=0.24\textwidth,trim=0.cm 1.2cm 0.cm 2.cm,clip]{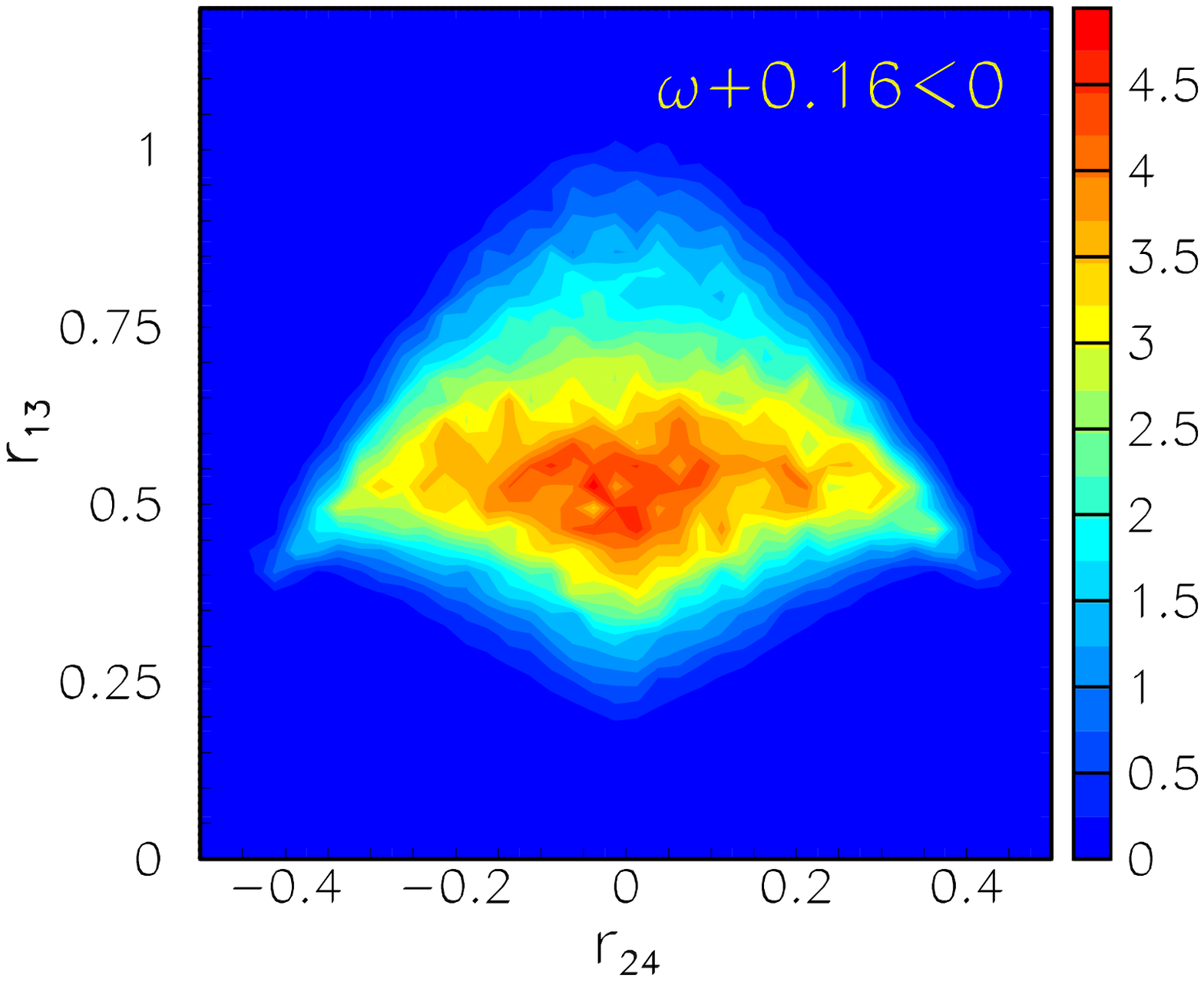}
\caption{Locations of the $\omega$+0.16$>$0 and respectively $<$0 in the $\theta_{24}$ vs $\theta_{13}$ plane (left panels) and $r_{13}$ vs $r_{24}$ plane (right panels).}
\label{fig9}
\end{figure*}
However,  the  criterion  used  in  Figure \ref{fig7}  to  define  long  and  short  axis  quads  does  not  differentiate between  them.  The  two  classes  are  well  separated  in  the $\eta_S$ vs $\xi_S$ plane  but we cannot tell  them  apart  from their location in the \mbox{$\theta_{24}$ vs $\theta_{13}$} or \mbox{$r_{13}$ vs $r_{24}$ planes}. This illustrates the limitations of using only two parameters in defining a quad configuration. Yet, we know that short axis and long axis quads can be told apart from the different parities of the 1$-$3 image pair: image 3 is inside the critical curve for long axis quads and outside for short axis quads, the opposite  being  true  for  the  2$-$4  pair. We translate this property by introducing a parameter $\omega$=$r_3-1/2(r_2+r_4)$ which is indeed observed \mbox{(Figure \ref{fig8})} to be a good discriminant between short-axis and long-axis quads. Its distribution separates the quads in two families, one associated with long axis and the other with short axis configurations. We estimate from Figure \ref{fig8} (left) that the separation between these occurs approximately at $\omega$$\sim$$-$0.16.  However, Figure \ref{fig9} shows that the two classes overlap in both the $\theta_{24}$ vs $\theta_{13}$ and \mbox{$r_{13}$ vs $r_{24}$ planes}. 
\begin{figure*}
  \centering
    \includegraphics[width=0.35\textwidth,trim=0.cm 1.2cm 0.cm 2.cm,clip]{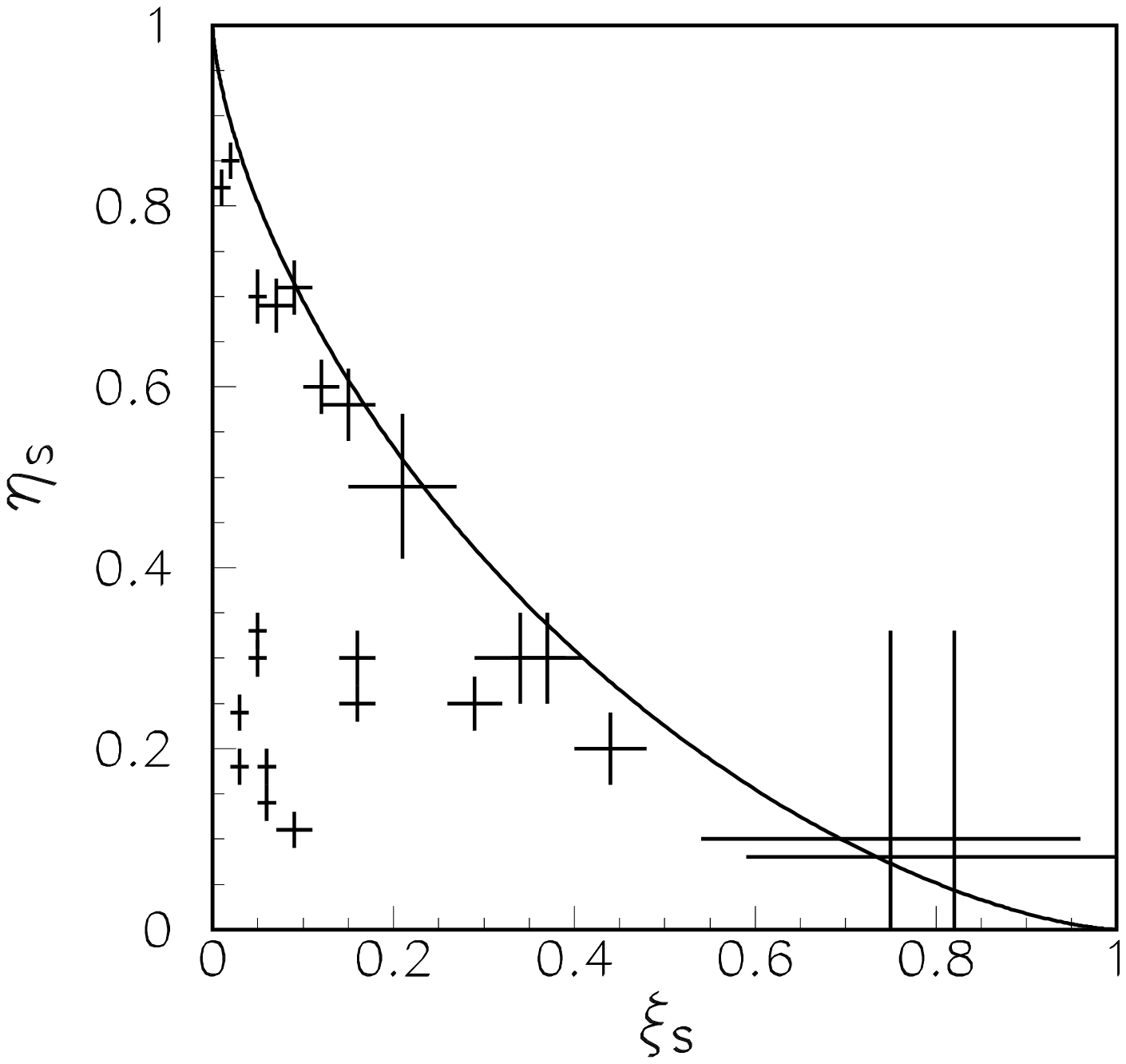}
    \includegraphics[width=0.35\textwidth,trim=0.cm 1.2cm 0.cm 2.cm,clip]{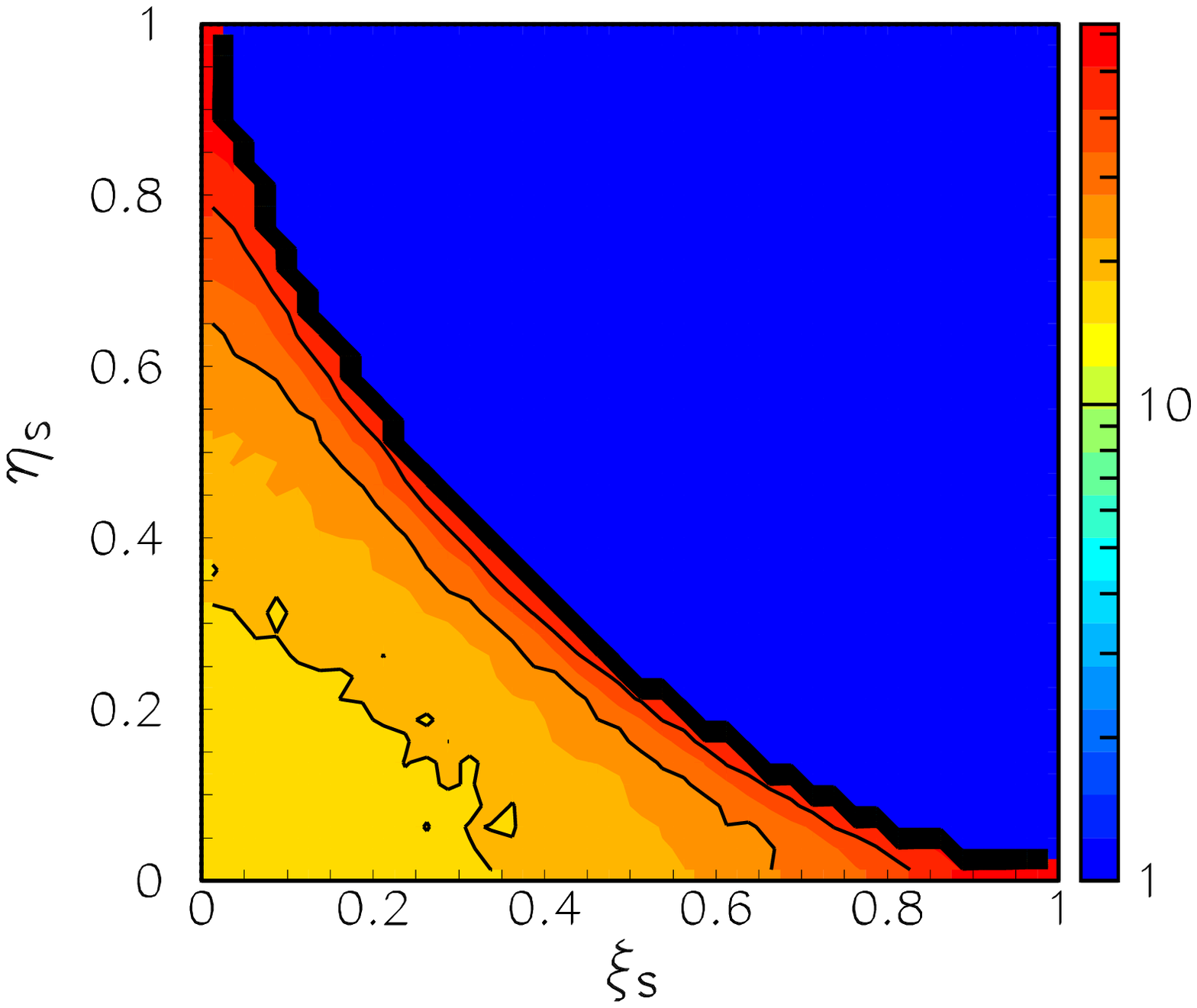}
\caption{Left:  the $\eta_S$ vs $\xi_S$ distribution  of the \citet{castles}  quad  sample  as  calculated  from  the ($\theta_{24}, \theta_{13}$) and ($r_{24}, r_{13}$) pairs (Table \ref{tab3}). The locations of RX J0911 and B2045+265 are evaluated from the $\theta$ pair exclusively (see text). Right: map of the mean total magnification (sum of the absolute values) in the $\eta_S$ vs $\xi_S$ plane.}
\label{fig10}
\end{figure*}
\begin{figure*}
  \centering
    \includegraphics[width=0.31\textwidth,trim=0.cm 1.2cm 0.cm 2.cm,clip]{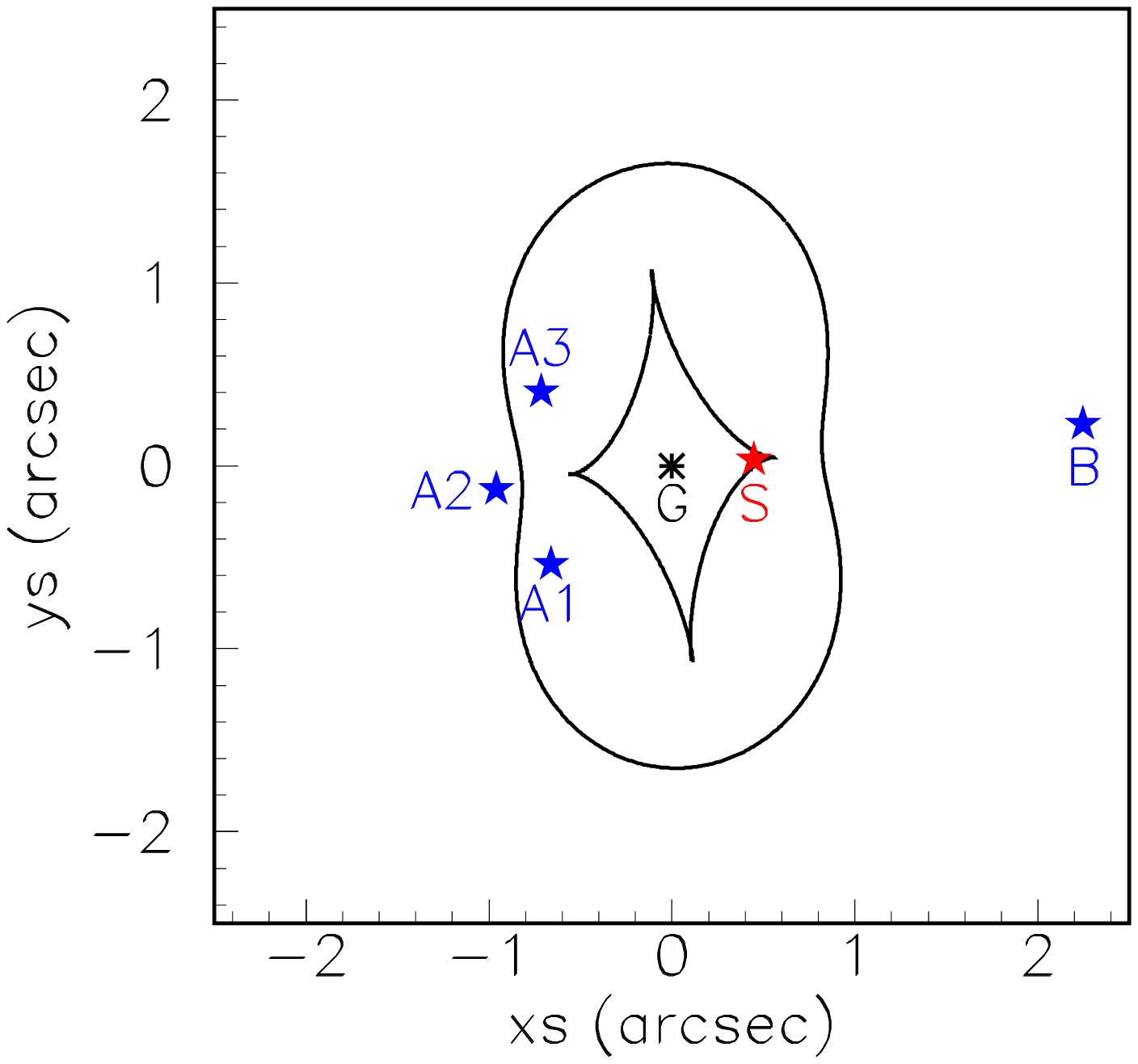}
    \includegraphics[width=0.31\textwidth,trim=0.cm 1.2cm 0.cm 2.cm,clip]{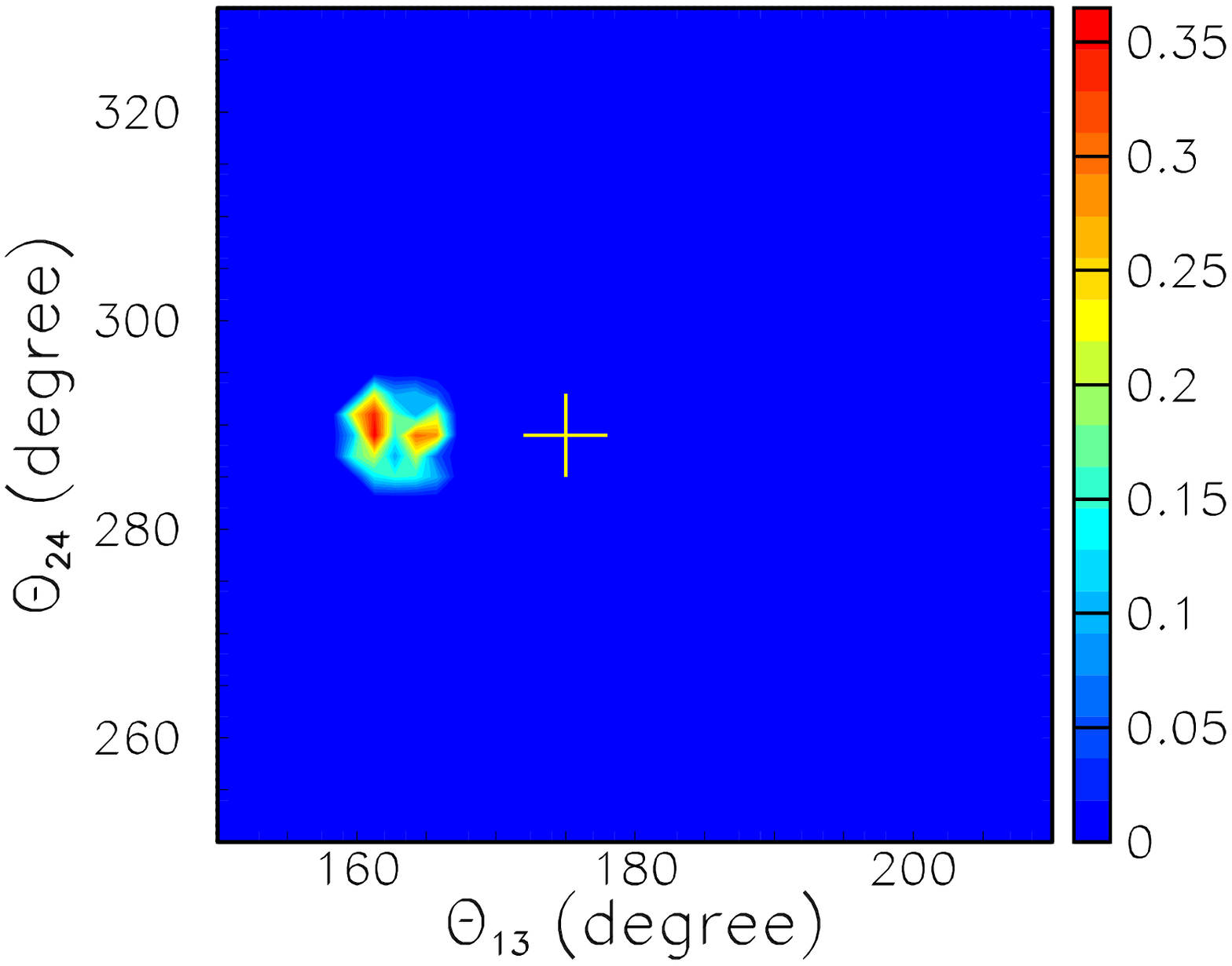}
    \includegraphics[width=0.31\textwidth,trim=0.cm 1.2cm 0.cm 2.cm,clip]{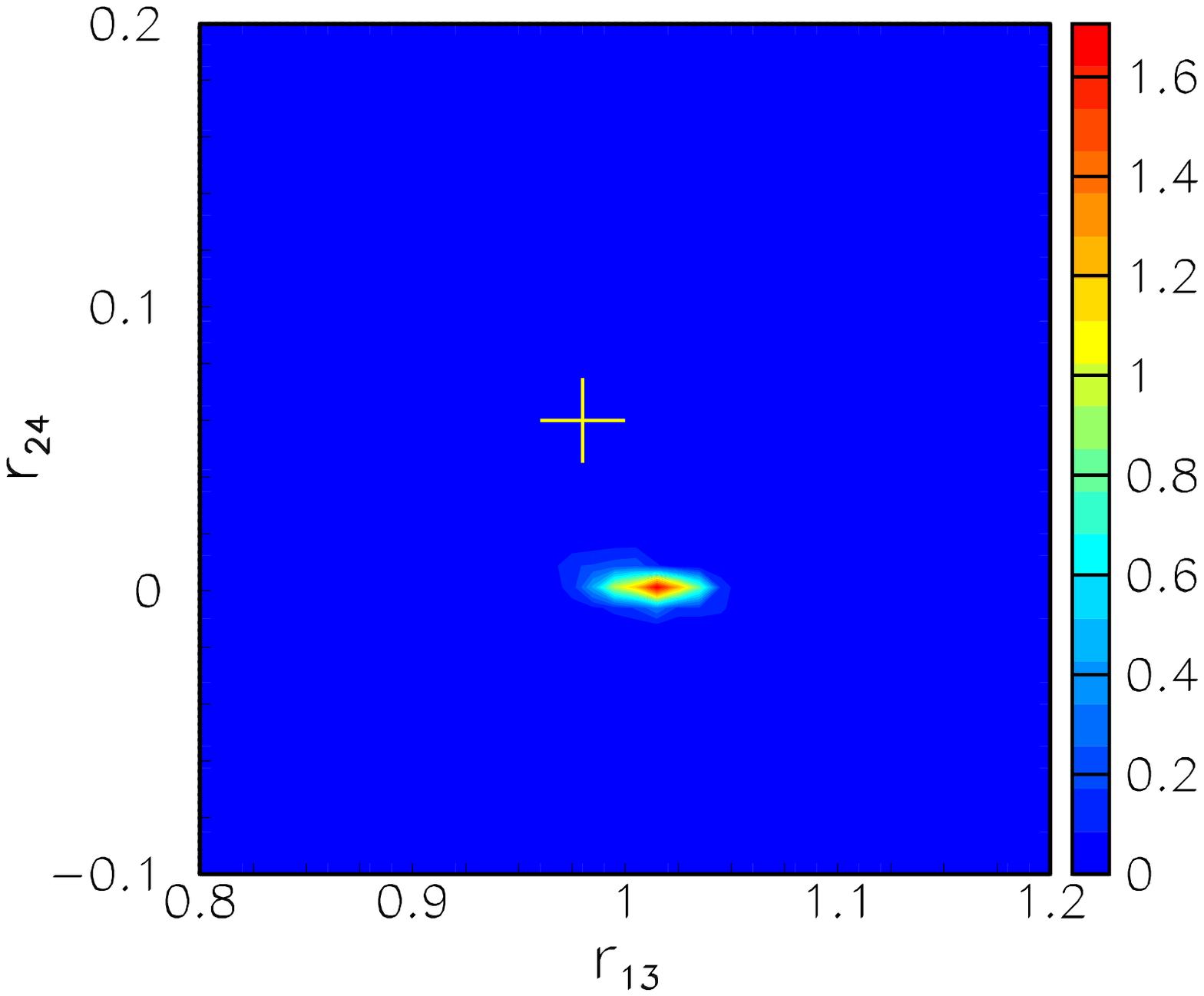}    
    \includegraphics[width=0.31\textwidth,trim=0.cm 1.2cm -1.cm 0.cm,clip]{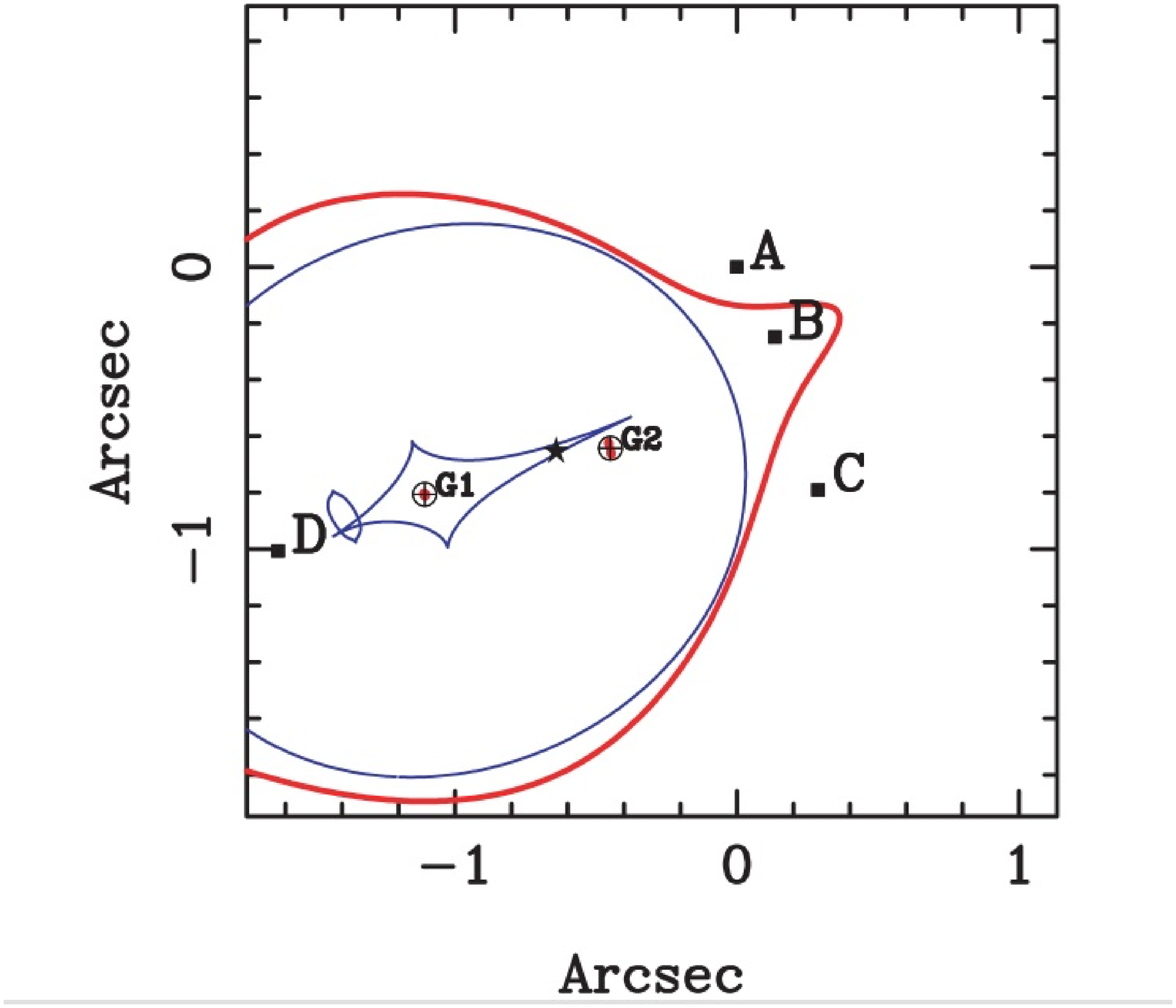}
    \includegraphics[width=0.31\textwidth,trim=0.cm 1.2cm 0.cm 2.cm,clip]{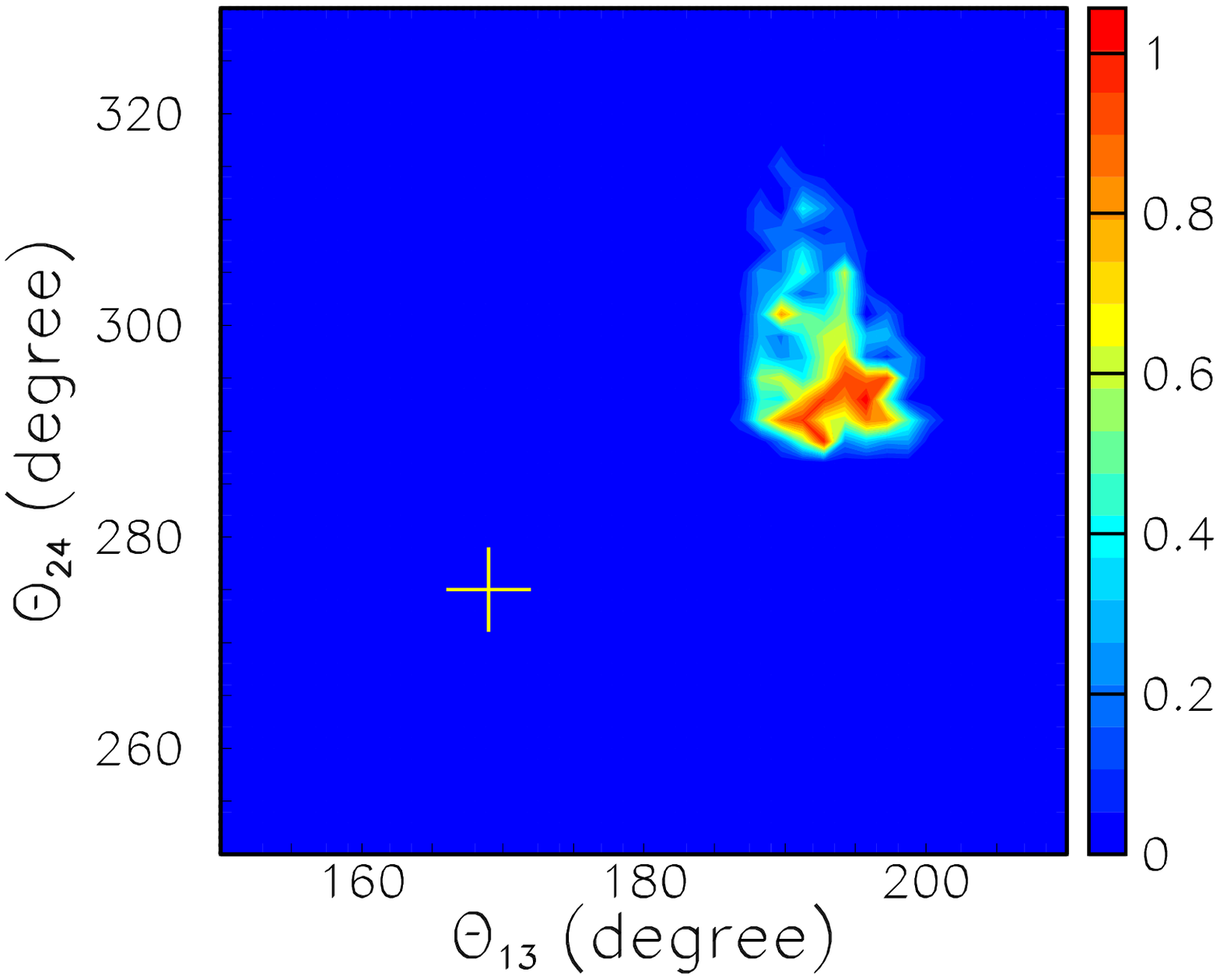}
    \includegraphics[width=0.31\textwidth,trim=0.cm 1.2cm 0.cm 2.cm,clip]{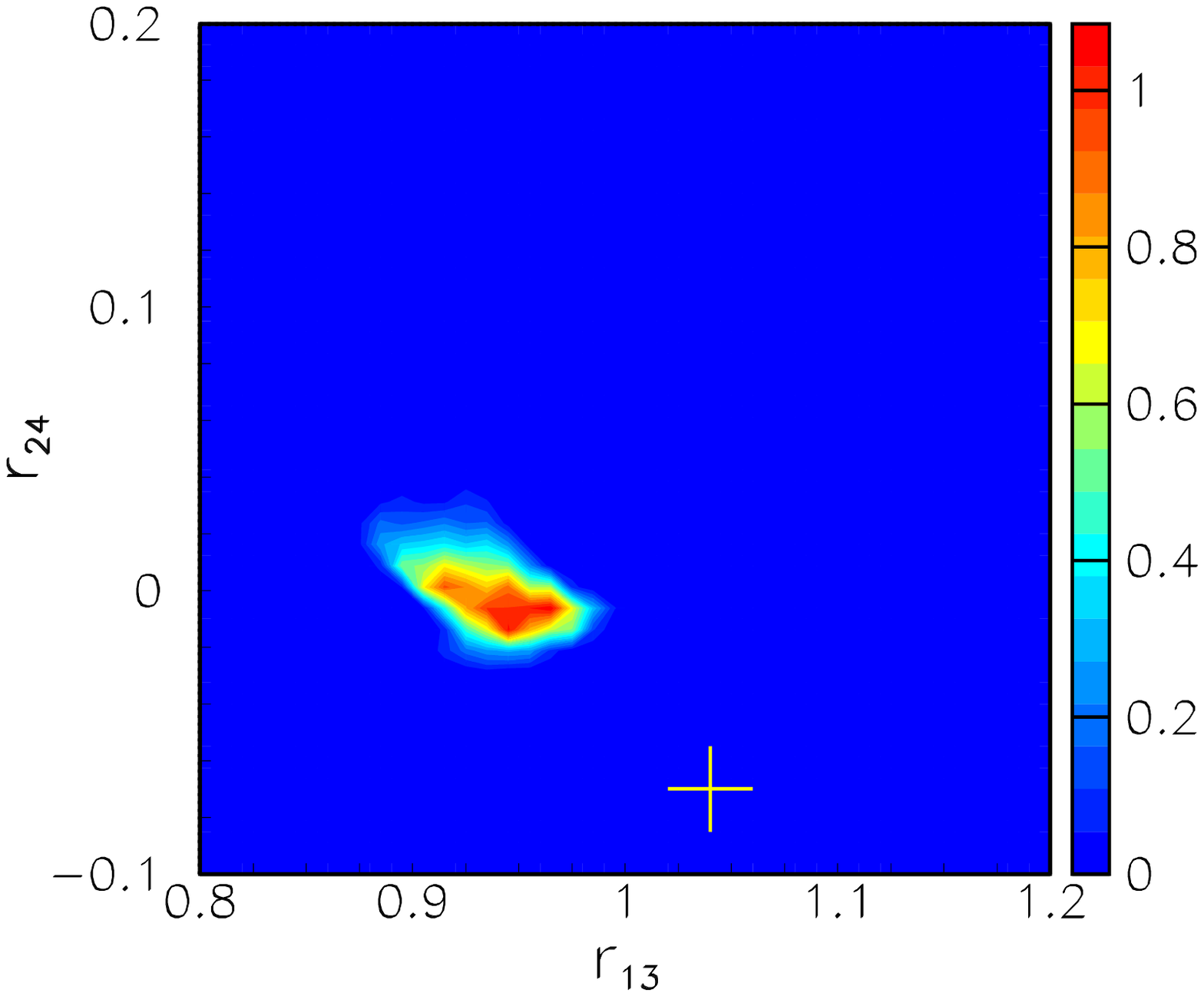}
\caption{The  cases  of  RXJ 0911 and  B2045+265  are  illustrated  in  the  upper and  lower  rows  respectively.  In  each row, from left to right, we display the results of appropriate lens models (\mbox{\citet{Hoai2013}} for RX J0911 and \citet{McKean2007} for B2045+265) and comparisons between prediction and measurement in the $\theta_{24}$ vs $\theta_{13}$ plane and in the $r_{24}$ vs $r_{13}$ plane  respectively. Predictions are  calculated  for $|r_{13}-0.98|$$<$0.02  and  $|r_{24}-0.06|$$<$0.02 (RX  J0911, $\theta_{24}$ vs $\theta_{13}$);  for \mbox{$|\theta_{24}-289\text{\dego}|$$<$5\dego and $|\theta_{13}-178\text{\dego}|$$<$5\dego(RX  J0911, $r_{24}$ vs $r_{13}$)};  for  $|r_{13}-1.04|$$<$0.05  and $|r_{24}+0.07|$$<$0.05 (B2045, $\theta_{24}$ vs $\theta_{13}$); for $|\theta_{24}-275\text{\dego}|$$<$5\dego and \mbox{$|\theta_{13}-169\text{\dego}|$$<$5\dego} (B2045, $r_{24}$ vs $r_{13}$). Crosses indicate the locations of the measured values.}
\label{fig11}
\end{figure*}
\subsection{Locating the quasar inside the caustic from the image configuration}\label{sec3.3}
\begin{table*}
  \centering
  \caption{Location of the quasar point source in the lens caustic obtained from the quad configuration and comparison of the associated model prediction with measurement.}
  \label{tab3}
  \begin{tabular}{c c c c c c c c }
    \hline\hline

    &{Name}
    &$\xi_s$
    &$Q(\xi_s)$
    &$\eta_s$
    &$Q(\eta_s)$
    &$Q(x_L)$
    &$Q(y_L)$\\
    \hline
    1
    &{PMNJ0134-0931}
    &0.02$\pm$0.01
    &0.01
    &0.85$\pm$0.02
    &0.22
    &$-$
    &$-$\\
   
    2
    &{HE0230-2130}
    &0.16$\pm$0.02
    &0.23
    &0.30$\pm$0.03
    &0.45
    &4.2
    &4.4\\
   
    3
    &{MG0414+0534}
    &0.09$\pm$0.02
    &0.36
    &0.71$\pm$0.03
    &0.53
    &1.9
    &4.0\\
   
    4
    &{HE0435-1223}
    &0.03$\pm$0.01
    &0.16
    &0.24$\pm$0.02
    &0.09
    &0
    &0.7\\
   
    5
    &{B0712+472}
    &0.07$\pm$0.02
    &0.05
    &0.69$\pm$0.03
    &0.27
    &$-$0.4
    &0.4\\
   
    6
    &{HS0810+2554}
    &0.12$\pm$0.02
    &0.55
    &0.60$\pm$0.03
    &0.67
    &2.8
    &$-$0.9\\
   
    8
    &{SDSS0924-0219}
    &0.29$\pm$0.03
    &0.29
    &0.25$\pm$0.03
    &0.04
    &$-$1.9
    &1.7\\
   
    9
    &{SDSS1004+4112}
    &0.21$\pm$0.06
    &0.07
    &0.49$\pm$0.08
    &0.09
    &0.3
    &$-$1.1\\
    
    10
    &{SDSS1011+0143}
    &0.06$\pm$0.01
    &0.66
    &0.18$\pm$0.02
    &0.67
    &0.9
    &3.9\\
   
    11
    &{PG1115+080}
    &0.37$\pm$0.04
    &0.18
    &0.30$\pm$0.05
    &0.27
    &$-$0.6
    &0.8\\
   
    12
    &{RXJ1131-1231}
    &0.01$\pm$0.01
    &0.10
    &0.82$\pm$0.02
    &0.33
    &0.5
    &$-$0.4\\
   
    13
    &{SDSS1138+0314}
    &0.16$\pm$0.02
    &0.25
    &0.25$\pm$0.02
    &0.45
    &$-$0.4
    &0.8\\
   
    14
    &{HST12531-2914}
    &0.05$\pm$0.01
    &0.64
    &0.33$\pm$0.02
    &0.14
    &0.7
    &0.1\\
  
    15
    &{HST14113+5211}
    &0.09$\pm$0.02
    &0.45
    &0.11$\pm$0.02
    &0.52
    &0
    &$-$2.1\\
  
    16
    &{H1413+117}
    &0.03$\pm$0.01
    &0.19
    &0.18$\pm$0.02
    &0.66
    &$-$2.7
    &0.3\\
  
    17
    &{HST14176+5226}
    &0.06$\pm$0.01
    &0.69
    &0.14$\pm$0.02
    &0.04
    &2.2
    &0\\
   
    18
    &{B1422+231}
    &0.05$\pm$0.01
    &0.18
    &0.70$\pm$0.03
    &0.45
    &$-$3.9
    &0.9\\
   
    19
    &{B1555+375}
    &0.15$\pm$0.03
    &0.02
    &0.58$\pm$0.04
    &0.07
    &2.6
    &0.4\\
  
    20
    &{WF12026-4536}
    &0.44$\pm$0.04
    &0.25
    &0.20$\pm$0.04
    &0.28
    &$-$1.8
    &1.1\\
   
    21
    &{WF12033-4723}
    &0.34$\pm$0.05
    &0.14
    &0.30$\pm$0.05
    &0.55
    &$-$0.3
    &$-$0.6\\
   
    23
    &{Q2237+030}
    &0.05$\pm$0.01
    &0.71
    &0.30$\pm$0.02
    &0.33
    &0.2
    &$-$0.1\\
    \hline\hline
  \end{tabular}
\end{table*}

The results of the previous Section show that it should be possible to evaluate the values of $\xi_S$ and $\eta_S$ from those of $\theta_{13}$ and $\theta_{24}$ or $r_{13}$ and $r_{24}$. To this end we use the model to produce maps of the mean and rms values of $\xi_S$ and $\eta_S$ in the $r_{13}$ vs $r_{24}$ and $\theta_{13}$ vs $\theta_{24}$ planes. However, in order to avoid difficulties with the ambiguous  cases illustrated in Figure \ref{fig8},  we separate the quads in two  families according to the sign of the quantity $\omega$+0.16. The  maps  of  the  mean and  rms values  are  displayed  in  Figure \ref{fig18}  of  the  Appendix for $\sigma_{\gamma}$=0.1. We  then  associate  to  each  \citet{castles} quad a  location in  the $\eta_S$ vs $\xi_S$ plane as obtained  from  these maps. Precisely, we obtain the mean and rms values of $\xi_S$ and $\eta_S$ for the $r$ pair and $\theta$ pair independently. Knowing $\theta_{13}$, $\theta_{24}$ and $\omega$, we obtain $<$$\xi_S$$>_{\theta}$, $<$$\eta_S$$>_{\theta}$, Rms($\xi_s$)$_\theta$ and Rms($\eta_s$)$_{\theta}$ from the appropriate map. Similarly, knowing $r_{13}$, $r_{24}$ and $\omega$, we obtain $<$$\xi_S$$>_{r}$, $<$$\eta_S$$>_{r}$, Rms($\xi_s$)$_{r}$ and Rms($\eta_s$)$_{r}$ from the appropriate map. We then combine the $r$-pair and $\theta$-pair evaluations as shown below. The result is listed in Table \ref{tab3} and illustrated in Figure \ref{fig10}. Table \ref{tab3} lists for each quad the mean values of the coordinates $\xi_S$ and $\eta_S$ of the source obtained from the ($\theta_{13}, \theta_{24}$) and ($r_{13}, r_{24}$) pairs, namely:
\begin{align*}
 \xi_S=&\frac{\{<\xi_S>_{\theta}/\mbox{Rms}^2(\xi_S)_{\theta}+<\xi_S>_r/\mbox{Rms}^2(\xi_S)_r \}}{1/\mbox{Rms}^2(\xi_S)_{\theta}+1/\mbox{Rms}^2(\xi_S)_r}\\&\pm \frac{1}{\left\{{1/\mbox{Rms}^2(\xi_S)_{\theta}+1/\mbox{Rms}^2(\xi_S)_r}\right\}^{1/2}}
\end{align*}
Similarly,
\begin{align*}
 \eta_S=&\frac{\{<\eta_S>_{\theta}/\mbox{Rms}^2(\eta_S)_{\theta}+<\eta_S>_r/\mbox{Rms}^2(\eta_S)_r \}}{1/\mbox{Rms}^2(\eta_S)_{\theta}+1/\mbox{Rms}^2(\eta_S)_r}\\&\pm \frac{1}{\left\{{1/\mbox{Rms}^2(\eta_S)_{\theta}+1/\mbox{Rms}^2(\eta_S)_r}\right\}^{1/2}}
\end{align*}
Here, $<$$\xi_S$$>_{\theta}$, $<$$\xi_S$$>_r$, $<$$\eta_S$$>_{\theta}$ and $<$$\eta_S$$>_r$ are the mean values of respectively $\xi_S$ and $\eta_S$ obtained from the maps of Figure \ref{fig18} for the $\theta$ and $r$ pair respectively.\\
Similarly, $\text{Rms}(\xi_S)_{\theta}$, $\text{Rms}(\xi_S)_r$, $\text{Rms}(\eta_S)_{\theta}$ and $\text{Rms}(\eta_S)_r$ are the Rms values of respectively $\xi_S$ and $\eta_S$ obtained from the maps of Figure \ref{fig18} for the $\theta$ and $r$ pair respectively.
For each quad we also list quality factors $Q(\xi_S)$ and $Q(\eta_S)$, which measure the agreement between the two evaluations (using the $\theta$ pair or the $r$ pair):
\begin{align*}
    Q(\xi_S)=\left\{\frac{(<\xi_S>_{\theta}-\xi_S)^2}{\mbox{Rms}^2(\xi_S)_{\theta}}+\frac{(<\xi_S>_r-\xi_S)^2}{\mbox{Rms}^2(\xi_S)_r}\right\}^{1/2}\\
    Q(\eta_S)=\left\{\frac{(<\eta_S>_{\theta}-\eta_S)^2}{\mbox{Rms}^2(\eta_S)_{\theta}}+\frac{(<\eta_S>_r-\eta_S)^2}{\mbox{Rms}^2(\eta_S)_r}\right\}^{1/2}
\end{align*}
The  agreement  is  excellent,  with $Q$ values  not  exceeding  0.7  and  having  an  average  of  0.3. In particular, it is remarkable that SDSS1004 (nr 9), in spite of being lensed by a galaxy cluster, is very well described by the simple lens model.  The quasar  point  sources  of  the  \citet{castles}  quad  sample  are  seen  to  populate  preferably  the region  close  to  the caustic boundary, where the magnification is important. This bias is well known. We illustrate it by mapping the  mean  magnification (in absolute value and summed  over  the  four  images) in  the $\eta_S$ vs $\xi_S$ plane  for model  quads  (Figure \ref{fig10} right).
\begin{figure*}
  \centering
    \includegraphics[width=0.35\textwidth,trim=0.cm 1.2cm 0.cm 2.cm,clip]{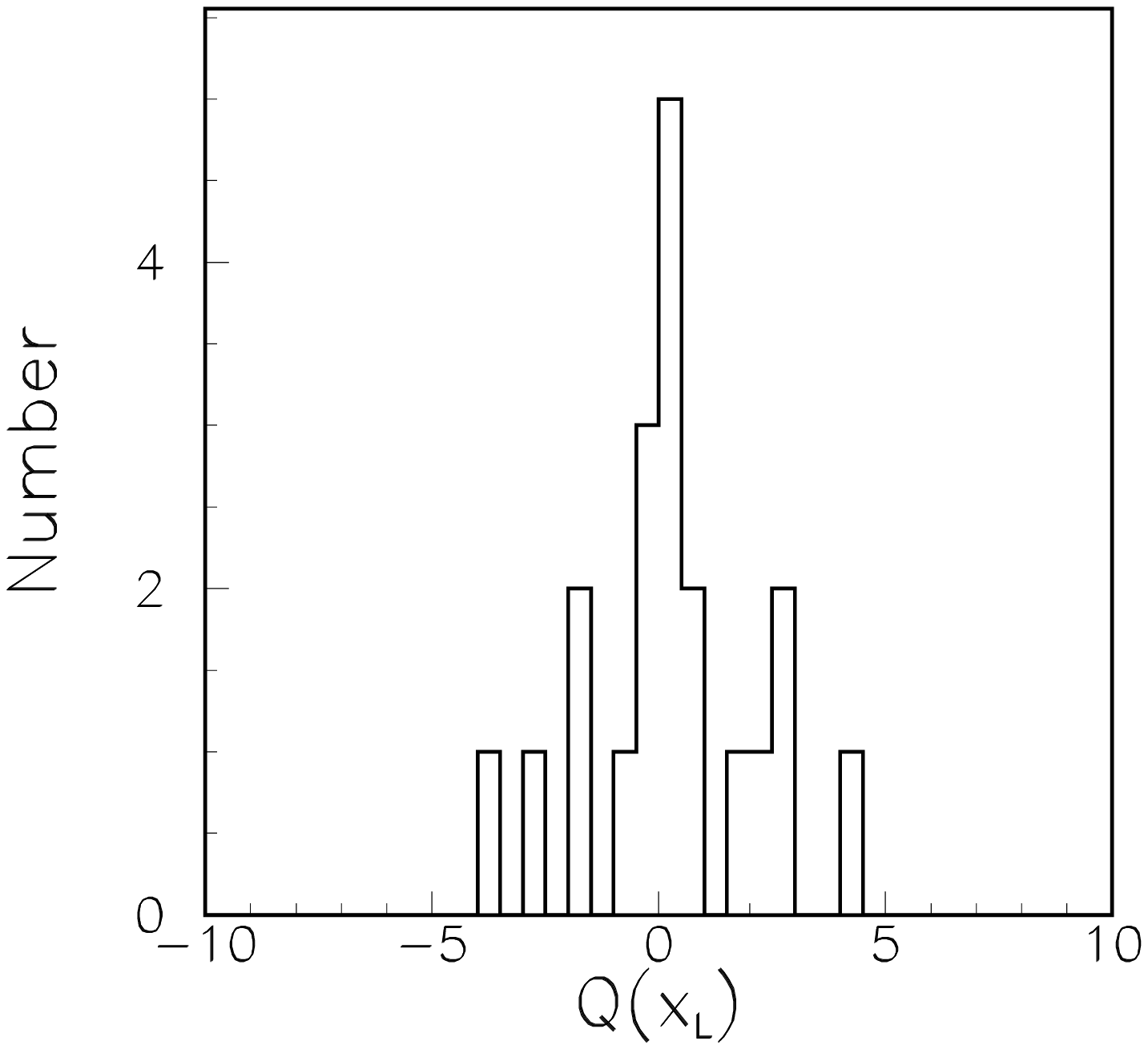}
    \includegraphics[width=0.35\textwidth,trim=0.cm 1.2cm 0.cm 2.cm,clip]{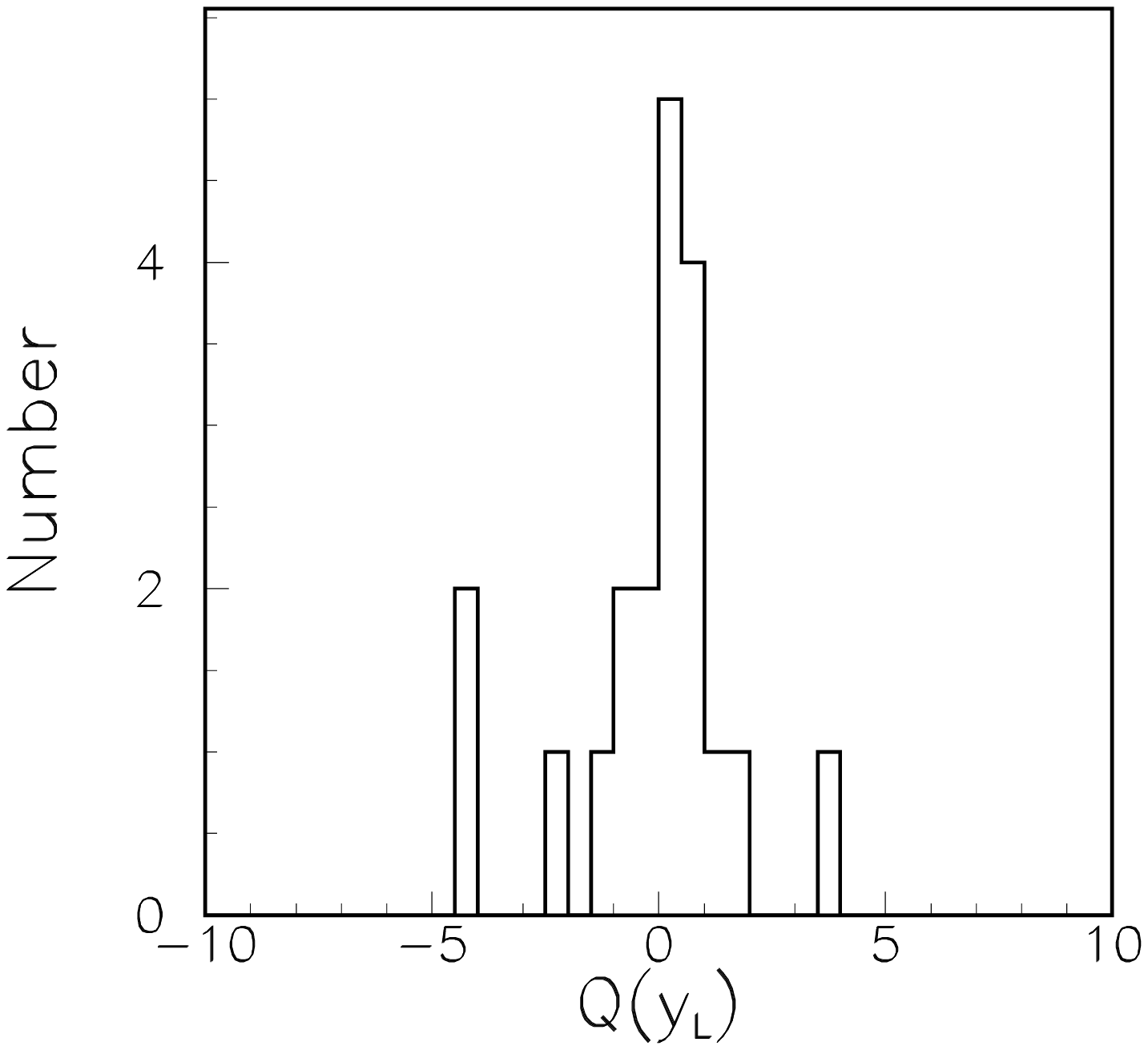}
\caption{Distribution of the quality factors $Q(x_L)$ (left) and $Q(y_L$) (right).}
\label{fig12}
\end{figure*}
Two \mbox{\citet{castles}} quads are omitted from Table \ref{tab3}: RX J0911 (nr 7) and \mbox{B2045+265 (nr 22)}. Both are short  axis  quads according  to  the  value  of $\omega$.  Their  values  of  ($\theta_{13}, \theta_{24}, r_{13}, r_{24}$)  are  \mbox{(178\dego,  289\dego,  0.98,  0.06)} and  \mbox{(169\dego, 275\dego,  1.04, $-$0.07)}  respectively,  making  them,  by  far, the quads  having  the largest  values  of \mbox{$r_{13}$ and $\theta_{24}$} \mbox{(see Figures \ref{fig5} and \ref{fig6})}. Their locations on the $r$ pair \mbox{maps of Figure \ref{fig18}} (fourth row) are very close to the  edge  of  the  acceptable  region.  In  such  cases  the  lens  equation  imposes  very  strong  constraints  on  the normalised  coordinates, as is  also  visible in Figures \ref{fig5}  and \ref{fig6}. The  point $\theta_{13}$=180\dego, $r_{24}$=0  is  a  centre  of symmetry, implying $\theta_2$=360\dego$-$ $\theta_4$.   In   its   vicinity there   is   very   little   room   to   accommodate   quad configurations. The  same  is  true  near  the  upper and  lower ends of  the $r_{13}$ vs $\theta_{24}$ plot. We  illustrate  this further in  \mbox{Figure  \ref{fig11}},  which displays  the  predicted  values of  the $r$ pair  when  the $\theta$ pair  is  set  equal  to  the measured  values and,  conversely,  the  predicted  values of  the $\theta$ pair  when  the $r$ pair  is  set  equal  to  the measured values. In both cases, the two sets are disconnected. This implies that the $r$ pair cannot be used to obtain  a  reliable  evaluation  of $\xi_S$ and $\eta_S$.  We  show  in  Figure \ref{fig10}  the  evaluations  obtained  from  the $\theta$ pair; however, they are associated with large error bars.
\subsection{The lens location: comparison with model prediction}\label{sec3.4}
The analyses presented in the preceding sections are independent of the location of the lens. They do not  require  that  it  be  observed.  In  order  to  study  the  additional  information  that such  an  observation provides, we call $x_L$ and $y_L$ the lens normalised coordinates, obtained from the measured values by applying the same coordinate transformation as for the images. Those of the \citet{castles} quad sample are listed in Table \ref{tab1} and compared with model prediction in the right panel of Figure \ref{fig5}. Note that changing the sign of $y_L$ leaves the configuration invariant: we can only  hope to measure $|y_L|$. In order to assess the ability of the model  to  predict  the  location  of  the  lens,  we  calculate  for  each  \citet{castles}  quad  the  values  of $x_L$ and $|y_L|$ predicted by the model when $\xi_S$ and $\eta_S$ take the values listed in Table \ref{tab3}. More precisely, we calculate their distributions  when $\xi_S$ and $\eta_S$ are  Gaussian  distributed  about  these  values  with  the  associated  dispersions listed in Table \ref{tab3}. We call $<$$x_L$$>$$^*$  and $<$$|y_L|$$>$$^*$ the means of these distributions and $\text{Rms}^*(x_L)$ and $\text{Rms}^*(|y_L|)$ their rms values. We compare these model predictions to the measured values $x_L$ and $|y_L|$ listed in Table \ref{tab1} by defining  two  quality  factors, \mbox{$Q(x_L)=(x_L-$$<$$x_L$$>$$^*)$/$\Delta(x_L)$}  and \mbox{$Q(y_L)=(|y_L|-$$<$$|y_L|$$>^*)$/$\Delta(|y_L|)$} where $\Delta(x_L)$ and $\Delta(|y_L|)$ are  obtained  by  summing  in  quadrature  an  estimated  measurement  error  of  0.02  with $\text{Rms}^*(x_L)$ and $\text{Rms}^*(|y_L|)$ respectively. Their  distributions  are  displayed  in  Figure \ref{fig12}. The mean$\pm$rms  values  are  0.3$\pm$1.9 and $-$0.2$\pm$1.8 respectively. A quad is  omitted  from  \mbox{Figure \ref{fig12}}: PMNJ0134$-$0931 (nr 1),  a  quasar for  which the \citet{castles} assignment of the lens is not reliable \mbox{\citep{Winn2002,Wiklind2018}}. The other cases show qualitative  agreement  between  measurement  and  model  prediction,  however nearly  \mbox{twice worse  than predicted.}

\section{SUMMARY}\label{sec4}
In line with the work of previous authors who gave evidence for simple considerations on the image configuration to offer ample information on the lensing mechanism without having recourse to a detailed modelling of the lensing potential, we have given a further illustration of such remarkable properties of gravitational optics. The  introduction  of  normalised  coordinates  aimed  at  defining  quad  configurations  independently from  location,  orientation  and  size  has  revealed  the  presence  of  strong  correlations  specific to the gravitational  lensing  mechanism. These  are  conveniently  expressed  as  correlations  between  the  opening angle of a pair of images of given parity and the radial difference of the pair of opposite parity. As a result a quad configuration can be approximately described by a pair of parameters: the $\theta$ pair, giving the values of the  opening  angles  of  the  odd  and  even  pairs of  images,  or  the $r$ pair,  \mbox{giving  the  values  of  their  radial differences.} Such description does not require the lens to be detected.

We  have  applied  these  considerations  to  the  study of  a  sample  of  HST  quads  collected  by  the \citet{castles}  collaboration  and  we  have  shown  how  the  location  of  the  quasar  point  source  within  the  lens caustic can be evaluated from the quad configuration. The agreement between the results obtained using the $\theta$ pair and  the $r$ pair  has  given  confidence  in  the  validity  of  the  description  of  the  lensing  mechanism  by  a very  simple  lens  model.  This  model  has  a  single  parameter,  the  amplitude  of  the  external  shear,  and  the results that have been obtained display very little dependence on its precise value.

The  study  has  shown  the  soundness  of  the  classification  proposed  by  \citet{Saha2003}  and quantified the relation that it implies between quad configuration and source location within the lens caustic. The special case of quads having the odd parity  images and the lens nearly  aligned has been  given special attention  and  the  constraints  imposed  on  their normalised coordinates,  related  to  symmetry  with  respect  to the line on which the lens and the odd parity images are located, have been commented upon. The distinction between long-axis and short-axis quads, which becomes trivially meaningless when the shear term, $\gamma$, cancels, has been found to be difficult to handle. We have introduced an ad hoc parameter, $\omega$, to this effect, constructed from a well known property of quad systems.

The \citet{castles} quad sample has been shown to be biased toward high magnification images, a result that is well known.  Our definition of normalised coordinates does not require the lens to be observed but, when available, the  measured  normalised  coordinates  of  the  lens  centre  have  been  found  in  general agreement with model predictions.

Image magnifications have not been discussed in the present article. They are the subject of many detailed studies triggered by the anomalies that are often observed, calling for appropriate interpretations (see for example \mbox{\cite{Keeton2003, Keeton2005}} and references therein). Addressing such questions is well beyond the scope of the present work; however, we note that the two cases that were singled out in the present study, RXJ 0911 and B2045+265, are part of the sample of anomalous flux quads identified by \mbox{\cite{Keeton2003, Keeton2005}}. But this is probably pure coincidence, the other members of the anomalous flux sample being evenly distributed among the CASTLeS quads.

In conclusion, this study has offered an interesting exploration of the general properties of quadruply imaged  quasars  and has deepened  our  understanding  of  the  lensing  mechanism  at  stake.  However,  its ambitions cannot  reach  further  than  that.  A serious  and  reliable  study of  gravitationally  lensed  images implies  the  construction  of  an appropriate lens  potential  and  the  resolution  of  the  associated  lens  equation, which nothing can replace. 
\acknowledgments
Financial  support  from  the  World  Laboratory  and  VNSC  is  gratefully  acknowledged.  This  research  is funded by the Vietnam National Foundation for Science and Technology Development (NAFOSTED) under grant number \mbox{103.99-2018.325.}
 \appendix
\section{}
\begin{figure*}[htb]
  \centering
    \includegraphics[width=0.8\textwidth,trim=0.cm .5cm 0.cm 0.cm,clip]{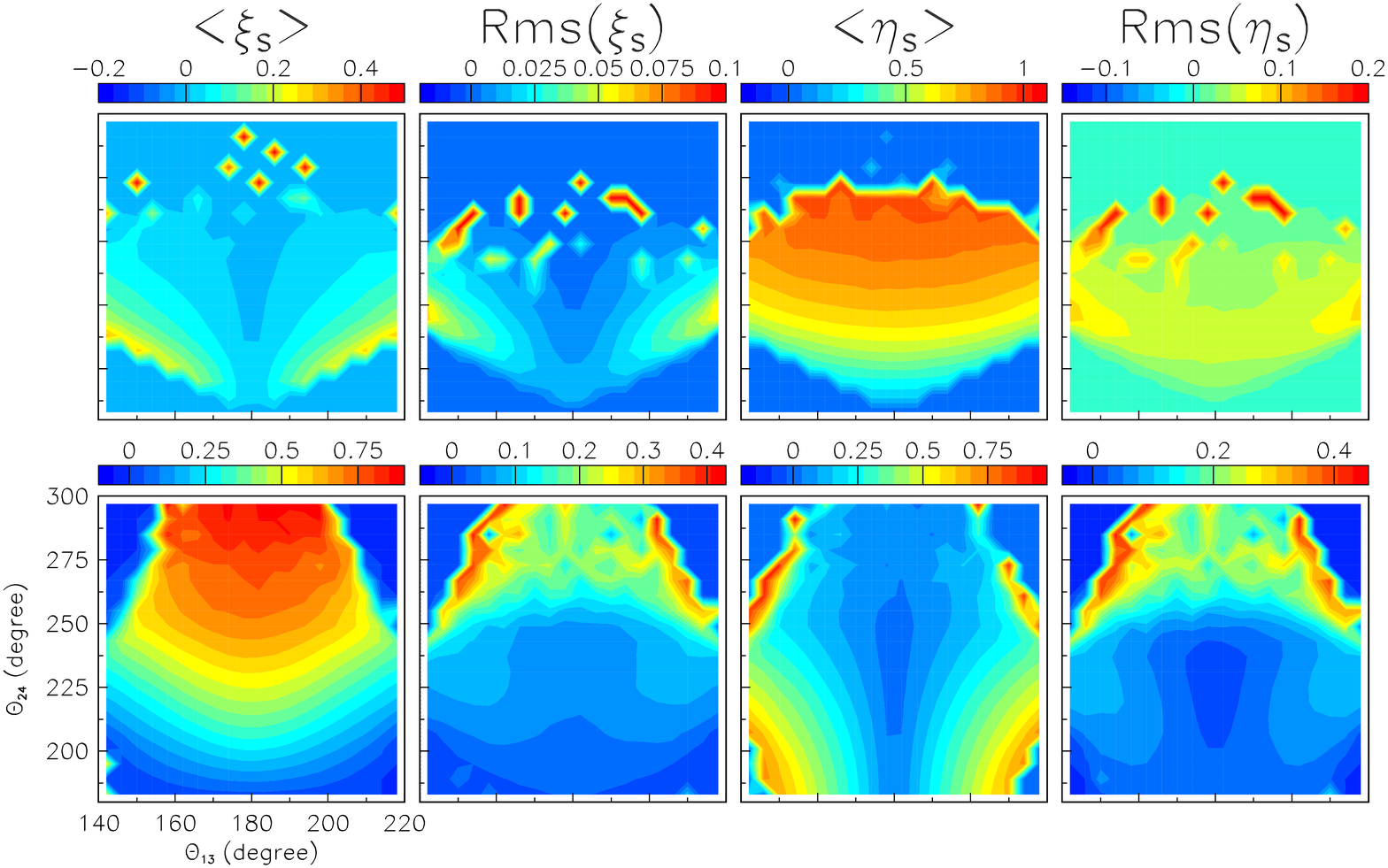}
    \includegraphics[width=0.8\textwidth,trim=0.cm .5cm 0.cm 0.cm,clip]{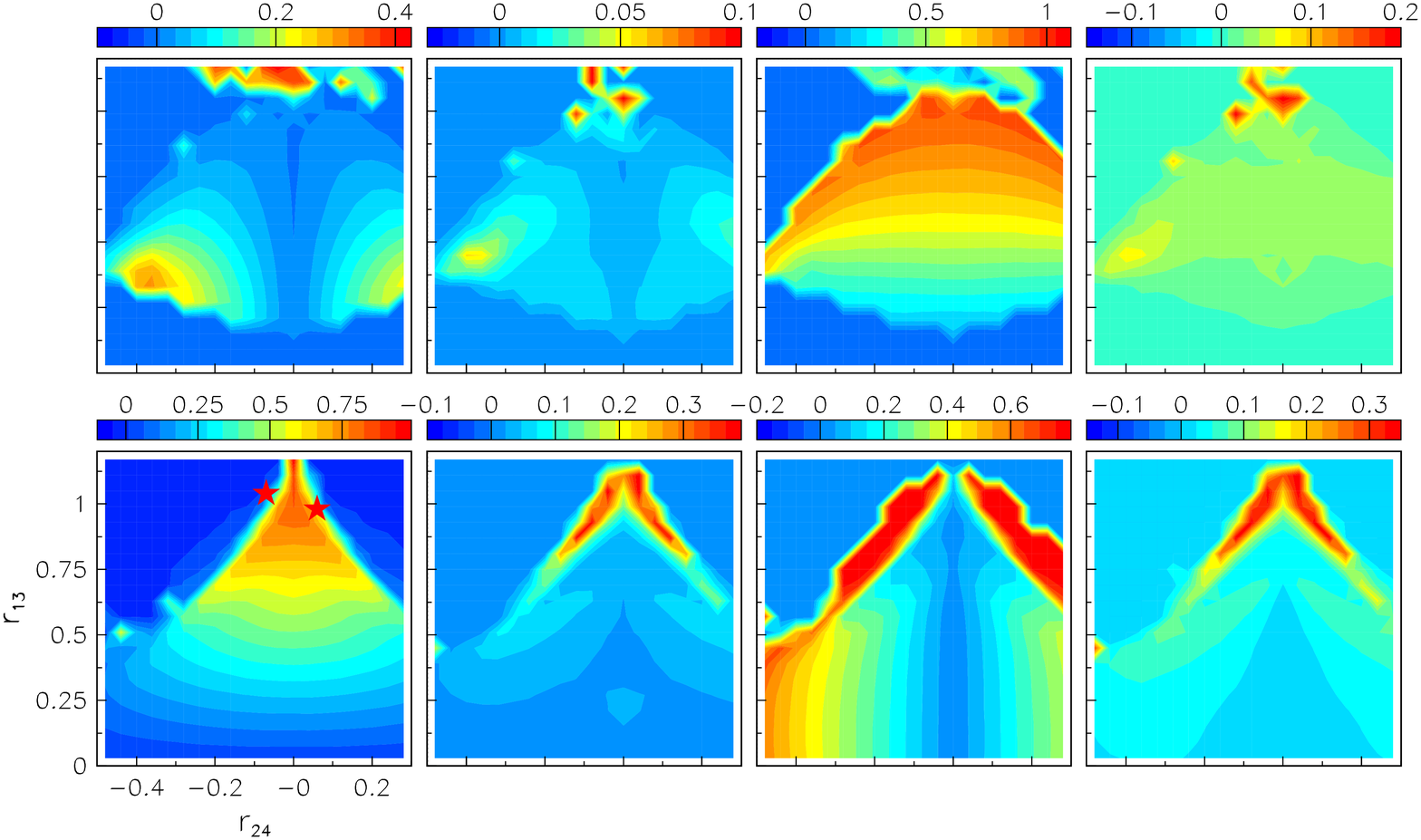}
\caption{Model maps of the mean values and rms values of $\xi_S$ and $\eta_S$ calculated for $\sigma_{\gamma}$=0.1, as indicated on top of each column. Rows go by pairs. The upper pair shows maps in the $\theta_{24}$ vs $\theta_{13}$ plane. The lower pair shows maps in the $r_{13}$ vs $r_{24}$ plane. In each pair the upper (lower) row is for $\omega$+0.16 negative (positive).}
\label{fig18}
\end{figure*}

\end{document}